\documentclass{article}
\usepackage{amssymb,multirow} 
\usepackage{amssymb}
\usepackage{cite,multirow} 
\usepackage[dvips]{graphicx,epsfig}
\usepackage{graphics}
\def\1ad{\mbox{\normalsize $^1$}}
\def\2ad{\mbox{\normalsize $^2$}}
\def\3ad{\mbox{\normalsize $^3$}}
\def\4ad{\mbox{\normalsize $^4$}}
\def\5ad{\mbox{\normalsize $^5$}}
\def\6ad{\mbox{\normalsize $^6$}}
\def\7ad{\mbox{\normalsize $^7$}}
\def\8ad{\mbox{\normalsize $^8$}}

\makeatletter
\@addtoreset{equation}{section}
\makeatother

\def\beq{\begin{equation}}                     %
\def\eeq{\end{equation}}                       %
\def\bea{\begin{eqnarray}}                     
\def\eea{\end{eqnarray}}                       
\def\0 {\nonumber}

\begin{document}

\setcounter{page}{0}
\begin{titlepage}
\titlepage
\rightline{SISSA-21-2006-EP}
\vskip 3cm
\centerline{{ \bf \Large Comments on the non-conformal gauge theories dual to $Y^{p,q}$ manifolds}}
\vskip .5cm
\vskip 1cm
\centerline{{\bf Andrea Brini$^*$ and Davide Forcella$^\dagger$}}
\vskip 1.5truecm
\begin{center}
\em 
 Mathematical Physics$^*$ and High Energy Physics$^\dagger$ sector \\
International School for Advanced Studies \\ 
Via Beirut 2, I-34014, Trieste, Italy\\
\vskip .4cm

\vskip 2.5cm
\end{center}
\begin{abstract}
We study the infrared behavior of the entire class of $Y^{p,q}$ quiver gauge theories. The dimer technology is exploited to discuss the duality cascades and support the general belief about a runaway behavior for the whole family. We argue that a baryonic classically flat direction is pushed to infinity by the appearance of ADS-like terms in the effective superpotential. We also study in some examples the IR regime for the $L^{a,b,c}$ class showing that the same situation might be reproduced in this more general case as well.

\vskip1cm

\end{abstract}
\vskip 0.5\baselineskip

\vfill
 \hrule width 5.cm
\vskip 2.mm
{\small
\noindent brini@sissa.it \\
forcella@sissa.it}
\begin{flushleft}
\end{flushleft}
\end{titlepage}
\large
\section{Introduction}

$D3$-branes living at the singularity of a $CY$ cone have provided general results for gauge/gravity correspondence since its early days.
The $IR$ limit of the gauge theory on the world volume of the $D3$-branes is dual to type $IIB$ string theory on the near horizon geometry $AdS_5 \times X$, where $X$ is the base of the cone: the horizon manifold \cite{horizon}. Since the cone is $CY$, $X$ is Sasaki-Einstein; the classical manifolds that admit a Sasaki-Einstein metric are $S^5$ (the original manifold of the correspondence), $T^{1,1}$ (the base of the conifold), and their quotients \cite{malda,kw}. \\
Recently, infinite classes of new Sasaki-Einstein metrics have been constructed \cite{gauntlett,CLPP,MSL,MS} and named $Y^{p,q}$ and $L^{a,b,c}$ and the corresponding dual gauge theories have been determined \cite{benvenuti,kru2,francolpqr,zaffa}. 
Thanks to recent developments in the correspondence \cite{intriligator,MSY} in particular when the $CY$ cone is a toric manifold \cite{Inv}, the explicit knowledge of the $CY$ metric is not so important for the construction of the dual gauge theories and for checks on the correspondence ($a$-maximization and $Z$-minimization) \cite{bertolini,kleb,kru,zaffa2,zaffa3}. In particular nowadays we have an algorithm that in principle allows us to construct the quiver gauge theories dual to all possible toric singularities \cite{hananyrombi,vafacorr}\footnote{For recent developments in this field see also \cite{sergiotri,seb,marvol}.}.
The interesting fact is that the correspondence can be extended to the non-conformal case. Indeed if there are some topologically non-trivial two cycles in the base of the $CY$ cones a simple way to break conformal invariance is to introduce fractional branes at their tip.     
Physically they correspond to $D5$-branes wrapped on two cycles collapsed at the singularity. At the level of the quiver, fractional branes correspond to rank assignments for nodes in the quiver, which are consistent with cancellation of non-abelian anomalies.
In the presence of a large number $N$ of $D3$-branes, a small amount $M$ of $D3$ fractional branes leads to a controlled breaking of conformal invariance, yielding field theories with tractable supergravity duals. 
In many cases, the field theory $RG$ flow is known to lead to a cascade of Seiberg dualities in which the effective number of $D3$-branes decreases, while that of fractional branes remains constant \cite{ks,kleb,hanany,hananyuranga2,strassler}. This flow tells us that the $IR$ dynamics is dominated by the field theories on the fractional branes, in absence of $D3$-branes.
Recently the supergravity solutions in the non-conformal case for the entire family of $Y^{p,q}$ and $L^{a,b,c}$ have been constructed \cite{kleb,gepner}.
Hence it is really interesting to understand the exact dynamics of the field theories on fractional branes.
In the cases of toric singularities it seems easy to classify the $IR$ dynamics of the gauge field theories:
just looking at the toric diagram we can understand the types of fractional branes that the geometry admits.
The number of possible fractional branes is equal to the number of non-trivial two cycles in the five dimensional manifold $X$. This is equal to the number of non-trivial three cycles that is the number of points on the boundary in the toric diagram minus three.
The types of fractional branes are divided into three classes related to the characteristic of the toric diagram\footnote{Usually a given geometry admits more than one type of fractional branes.} \cite{hananyuranga}:
\begin{itemize}
\item{toric diagrams with points on the boundary of the polygon $\Rightarrow$ $\mathcal{N}=2$ fractional branes: the gauge field theories on these branes have enhanced supersymmetry}
\item{toric diagrams that can be decomposed in Minkowski sums of polytopes $\Rightarrow$ deformation fractional branes: the gauge field theories on these branes confine in the $IR$}
\item{diagrams without points on the boundary or complex deformations, or generic composition of the previous two categories of fractional branes $\Rightarrow$ supersymmetry breaking ($SB$) fractional branes: the gauge theories on these branes break supersymmetry}
\end{itemize}  
One interesting class of such dynamics is given by the deformation fractional branes \cite{hananyuranga2, pin, hurdef, ago}. 
Geometries dual to these branes admit deformations and may allow the construction of a supergravity solution of the $KS$ type \cite{ks}. Some particular classes of these type of geometries are given by the generalized conifold and in particular by the $SPP$ singularity and it is an interesting subject to try to deform the singular metric for the construction of a complete regular supergravity solution.\\    
Clearly the most interesting one also for phenomenological reasons is the last type. The possibility of a gauge theory, with gravity dual,  which at the end of its flow in the $IR$ breaks dynamically the supersymmetry is certainly an interesting subject. The point is that the distinction between a stable non supersymmetric vacuum and runaway behaviour is not clear from the geometry.
It seems that branes related to the third class of fractional branes break in some sense supersymmetry, but it is not clear whether supersymmetry is in some way restored at infinity or there can be any stable non-supersymmetric vacuum.
Some cases were recently discussed in \cite{berenstein,hananyuranga,bertolini2} and it seems that the general behaviour is runaway \cite{hananyuranga}. The complete calculation of the $IR$ dynamics for the specific case of $Y^{2,1}$ theory was done in \cite{seiberg} confirming the general belief.\\
The proposal of this work is to give arguments in favour of this runaway behaviour for the complete class of $Y^{p,q}$ and some examples of $L^{a,b,c}$ singularities. We will work in the framework of field theories with gauge group given by a product of $SU(N_i)$ factors, because we consider the theories on the $D$-brane at low energy, where the $U(1)$ factors are massive and thus decoupled. Our aim is to show that in the cases under consideration it is always possible to find a gauge-invariant baryonic classically flat direction containing elementary fields which are sent to infinity by non-perturbative effects. This automatically solves the $D$-term equations of the field theory \cite{TJ} and gives hints towards the existence of runaway behavior for all these theories. Nevertheless, the possibility that some effects from the non-chiral sector of the action (i.e., the K\"ahler potential) may lead to a stabilization of this runaway mode is still left open, and is beyond the scope of this paper.\\
The paper is organized as follows. In section \ref{GeneralYpq} we give a brief review of $Y^{p,q}$ geometry and the  dual gauge theory. Here we explain the structure of the theory in the conformal and the non-conformal regime. In section \ref{Cascades} we discuss how to perform Seiberg dualities in the framework of the dimer technology. We describe the general structure of the $RG$ flow for $Y^{p,q}$ theories and we recast the known results on cascades for $Y^{p,p-1}$ and $Y^{p,1}$ in the dimer language; further, we construct the periodic structure of the coupling constant flow for these two subclasses. Moreover, some self-similar pattern of Seiberg dualities for the general case is explicitly derived: we point out how the correspondence between symmetries of the dimer and periodicity of the coupling constant flow under Seiberg duality appears to be more difficult to find in full generality. In section \ref{DSBrun} we give a short review of the $DSB$ and runaway behavior for these classes of gauge theories dual to toric cone. We discuss also some examples known in literature of quiver gauge theories with infrared ``$SB$''. In section \ref{Ypqdimer} we start the analysis of the low energy structure for the whole $Y^{p,q}$ family, finding out the general pattern for the appearance of non-perturbative Affleck-Dine-Seiberg terms in the effective superpotential from the study of the brane tiling representation of these theories. This is further carried over in section \ref{YpqIR}, where the analysis of the $F$-terms suggests that the behavior of the entire $Y^{p,q}$ class is runaway. In fact we find that it is always possible to construct a classically flat baryonic direction that is lifted to a runaway direction by non-perturbative effects. Finally, in section \ref{BeyondYpq} we outline how this probable scenario may be likely reproduced for $L^{a,b,c}$ dual gauge theories using some examples. Clearly here the family is huge and really complicated, and something different may happen. We hope to return to the general case of dimer gauge theories and toric geometry correspondence in the class of $SB$ fractional branes in future work.

\section{$Y^{p,q}$ quiver gauge theories}
\label{GeneralYpq}
\subsection{The geometry side}
\begin{figure}[!h]
\centering
\includegraphics[scale=0.5]{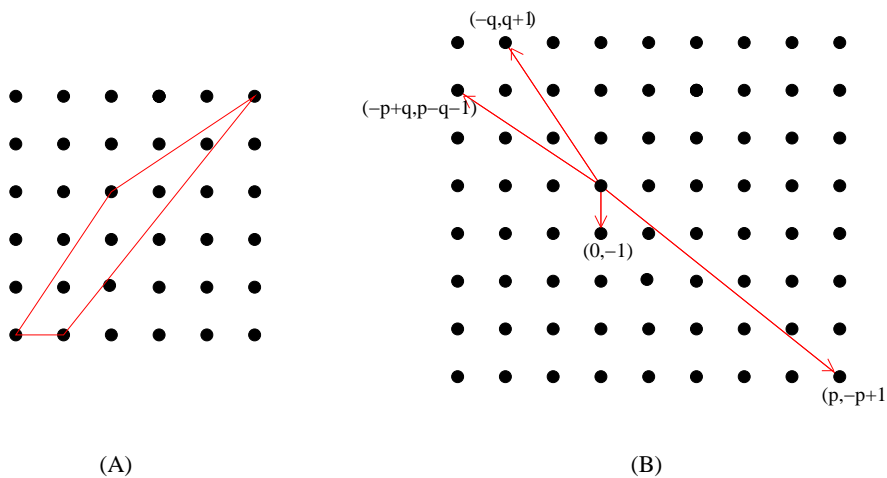}
\caption{(A) The toric diagram of $Y^{p,q}$ manifolds. (B) The $(p,q)$-wed diagram of $Y^{p,q}$ manifolds. In the figure we show the specific case of $Y^{5,2}$}
\label{toricy52}
\end{figure}

The $Y^{p,q}$ manifolds are an infinite class of five dimensional Sasaki-Einstein manifolds parameterized by two integers $p,q$ that satisfy the relation\footnote{We are only interested in the regular cases. We will neglect the possibility of having $Y^{p,0}$, $Y^{p,p}$ types of manifolds; the first ones correspond to $Z_p$ quotients of the conifold and admit complex deformations that are dual to confining gauge theories, while the second ones are orbifolds of flat space and the dual non-conformal gauge theories will be of $\mathcal{N}=2$ type.} $p > q > 0$. 
All these spaces have topology $S^3 \times S^2 $ and their metrics have $SU(2)\times U(1) \times U(1)$ isometry.
Our starting point is to consider a real cone over these spaces $C(Y^{p,q})$. These are toric $CY$ cones whose algebraic properties can be described by a toric diagram \cite{fulton} like in figure \ref{toricy52}.
The $CY$ condition tells us that the end points of the vectors in the lattice cone are all on a plane. This means that we can describe the geometry simply drawing a convex polygon on a plane.
The generic algebraic singularity is given in fact by a four points toric diagram, with vertices given by: 
\begin{equation}
\label{td}
v_1 = (p-q-1,p-q) \quad v_2 = (1,0) \quad v_3 = (p,p) \quad v_4 =(0,0)
\end{equation}
In this paper we will be interested in the study of the dual gauge theories that have no supersymmetric vacuum. The general feeling is that these theories are dual to geometries without complex deformations. 
The easiest way to understand if a toric singularity has complex deformation or not is to draw the $(p,q)$-web (dual of the toric diagram)  and see if it is possible to split it into subwebs in equilibrium \cite{altmann}, that are subwebs made of a set of vectors whose sum is equal to zero (see also \cite{hananyuranga2} and references therein).
This is clearly not possible for the case in figure \ref{toricy52}; hence, they do not admit deformation fractional branes. Moreover, the generic toric diagram has no lattice points on the edges, and therefore the geometry does not admit $\mathcal{N}=2$ fractional branes either. For these reasons they admit only ``$SB$'' fractional branes.

\subsection{The dual gauge theory}

\begin{figure}[!t]
\centering
\includegraphics[scale=0.5]{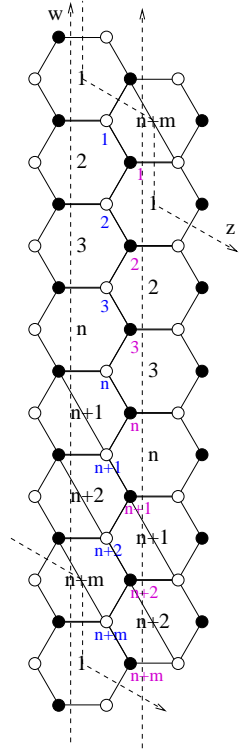}
\caption{The generic dimer for $Y^{p,q}$}
\label{kast}
\end{figure}

Now that we have understood the geometry, it is important to explain the dual gauge theory.
To construct the gauge theory we put a stack of $D3$-branes on the toric singularity transverse to the six dimensional $CY$ geometry.
The gauge theories which live on the worldvolume of the branes will be a superconformal $\mathcal{N}=1$ quiver gauge theory whose peculiar characteristics are described by a dimer diagram \cite{boxes,boxes2,kenyonvafa,dimers}. This is a set of white and black points together with a set of links which stretch only between points of different colours, drawn on a $T^2$ torus. This graph realizes a tiling of the torus in which every face corresponds to a gauge group, every link to a bifundamental chiral superfield which tranforms following the orientation given on the graph by the pattern of black and white points, and every point corresponds to a term in the superpotential which contains the chiral fields attached to the point and with the plus sign if the vertex is white and the minus sign if the vertex is black. Note that the superpotential is a sum of positive and negative terms, with each field appearing exactly once in a positive term and once in a negative term. This is reflected by the fact that the vertices of the dimer are colored in black or white and links connect only vertices of different color.
Nowadays there exists an algorithm that allows to construct the dual quiver gauge theory starting from any toric singularity \cite{hananyrombi}.
The quiver gauge theories were originally derived in \cite{benvenuti}, but we will use the language of dimers and we will present the theories following the convention of \cite{zaffa}.
The gauge theories on $Y^{p,q}$ have dimers built with $n$ hexagons and $2\,m$ quadrilaterals, that can be obtained by dividing into two parts $m$ hexagons drawn below the first $n$ ones (see figure \ref{kast}). There is only one column of such hexagons and divided hexagons that are periodically identified in the vertical direction. The other $T^2$ periodicity is described by drawing other columns and identifying each face with the face in the right column shifted down by one position $(k=1)$.    
The geometry is completely characterized by a pair of integer $(p,q)$, while the gauge theory is given in terms of another pair of integers $(n,m)$.
These two pairs are related by the following equations:
\begin{equation}
\left\{
\begin{array}{l}
n\,=\,2q\\
m\,=\,p-q
\end{array}
\right.
\end{equation}
The numbers of gauge factors, terms in the superpotential and  fields are given by
\begin{equation}
\begin{array}{r@{\,\,=\,\,}c@{\,\,=\,\,}l}
\# \,\textrm{gauge} & 2\,p & n+2\,m\\[0.5em]
\# \, \textrm{super potential} & 2\,p + 2q & 2\,(n+m)\\[0.5em]
\# \, \textrm{fields} & 4\,p + 2\,q & 3\,n+4\,m\\
\end{array}
\end{equation}
We are now in the conformal regime; if we apply Seiberg dualities to these quiver gauge theories we can obtain different quivers that flow in the $IR$ to the same $CFT$. It turns out that one can always find phases where all the gauge group factors have the same number of colors $N$; these are called toric phases.
Inside these toric phases we can find phases in which the number of chiral fields is equal to what we have written above. These special phases are called minimal toric phases, because they are the ones with a minimal number of fields. We shall consider here the theories only in minimal toric phases.
In these cases the gauge group is:
\begin{equation}
\prod_i^{2p} SU(N)
\end{equation}
These theories have one $SU(2)$ global symmetry, one $U(1)$ $R$-symmetry and one flavour $U(1)$ symmetry corresponding to the isometries of the $Y^{p,q}$ manifolds. 
The global symmetries in the gauge theory correspond on the supergravity side to massless gauge vectors in $AdS$ obtained by $KK$ compactification on the compact submanifolds $Y^{p,q}$. Inspecting the massless vectors we can find another $U(1)$ factor. This is the baryonic symmetry and the corresponding gauge field comes from the reduction of the $RR$ four form field of type $IIB$ supergravity on the non-trivial $S^3$ in $Y^{p,q}$. 
We can divide the chiral superfields in the theories in four classes:
\begin{itemize}
\item{$p-q$ singlets $Z$ fields.}
\item{$p+q$ singlets $Y$ fields.}
\item{$p$ doublets $U^{\alpha}$ fields.}
\item{$q$ doublets $V^{\alpha}$ fields.}
\end{itemize}
For us the important characteristic that distinguishes the four classes will be the baryonic charge (see table \ref{bcharges}).
\begin{table}[h!!!!!!]
\begin{center}
\begin{tabular}{c|c|c|c|c} \hline 
\, & $Z$ & $Y$ & $U$ & $V$ \\[0.5em] \hline 
$U(1)_B$ & $p+q$ & $p-q$ & $-p$ & $q$\\[0.5em] \hline 
\end{tabular}
\end{center}
\caption{Charges under the baryonic symmetry.}
\label{bcharges}
\end{table}
Since in this paper our principal task is to describe the $IR$ dynamics of the non-conformal theories dual to $Y^{p,q}$ manifolds we have to find a way to escape from the conformal case. 
The classical way to accomplish this is to add to our theories a number $M$ of fractional branes. These correspond to $D5$-branes wrapped on topologically non-trivial two cycles in the geometry that have vanishing volume.
Since the topology of smooth $Y^{p,q}$ is $S^3\times S^2$ we expect the
existence of a single type of fractional brane we can add in order
to obtain non-conformal theories. These $D5$-branes are wrapped
on the $S^2$ which is vanishing at the tip of the cone. The addition of
$M$ fractional branes gives a theory with the same number of gauge groups,
but with different number of colors
\begin{equation}
\prod_i^{2p} SU(N_i)
\end{equation}
and the same bi-fundamental fields as before. The numbers $N_i$
correspond to the unique assignment of gauge groups that lead to a 
non-anomalous theory. These numbers can be easily determined from 
the baryonic symmetry: the difference between the number of colors of
gauge groups connected by a bi-fundamental is $M$ times the baryonic charge
of the chiral field, with the convention that fields of opposite chirality
contribute with opposite sign. As a practical recipe, one can start
with a face in the tiling, assign to it
a conventional number of colors $N$, find the numbers in the adjacent
faces using the previous rule and then continue until all numbers $N_i$
are determined. We will see some examples of the realization of this procedure in the next sections.
The method works for the following reasons
\cite{kaza,hananyuranga}. First of
all, the cancellation of all anomalies for the baryonic symmetry implies
that the sum of charges for all the bi-fundamental fields starting from
a given face must vanish
\begin{equation}
{\rm face}\,  k:\qquad\qquad \sum_{i_k} q_{i_k} =0
\end{equation}
This condition ensures that, with the previous assignment, all gauge groups
are not anomalous. In addition, the fact that the baryonic symmetry is
a combination of the $U(1)$ factors in $\prod_i^{2p} U(N_i)$
guarantees that the result is independent on the choice of path on the
tiling used for determining the $N_i$'s.

\section{Cascades}
\label{Cascades}
As explained in the previous section the addition of a well behaved set of fractional branes breaks conformal symmetry. In fact, as we have seen, in the field theory side this corresponds to change the rank of some of the gauge group factors and it is easy to see that this operation upsets the vanishing of the beta functions \cite{gups,nek}. In this situation we clearly have a non-trivial $RG$ flow that is generically realized as duality cascade \cite{ks,kleb,hananyuranga2,strassler}.
In this section we want to study the $RG$ flow for the non-conformal gauge theories dual to the $Y^{p,q}$ manifolds in presence of fractional branes, deriving the general formulae for the beta functions and analyzing the cascades.

\subsection{$\beta$ functions}

\begin{figure}[h!]
\centering
\includegraphics[scale=0.6]{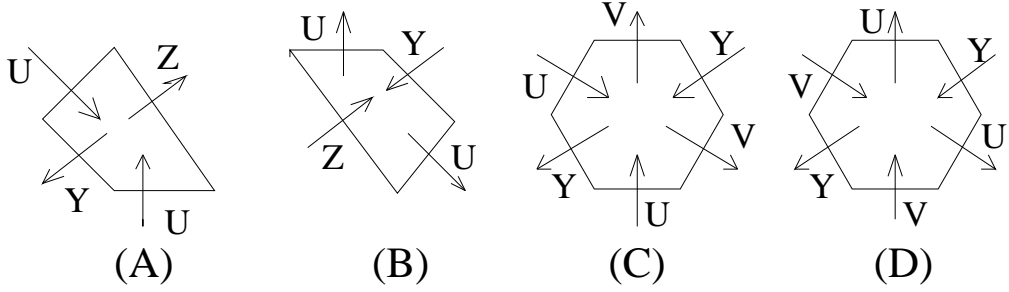}
\caption{The four different families of gauge factors that give the four different beta functions in the theory.}\label{beta}
\end{figure}

The $\beta$ function of the $i$-th gauge factor of the theory is \cite{shifman} 
\begin{equation}\label{betaf}
\beta ( g ) = \frac{ \partial g ( \mu ) }{ \partial \ln (\mu)} = - [3 T(\textrm{adj}) - \sum_i T(r_i)(1 - \gamma _i )] \frac{g^3}{16 \pi ^2}=- H \frac{g^3}{16 \pi ^2}
\end{equation}
where the sum runs over all the chiral fields trasforming in the $r_i$ representation of the gauge group, and having anomalous dimension $\gamma _i$. $T(r_i)$ is the Casimir of the $r_i$ representation normalized such that $T(\textrm{adj})= N_c $ and $T(\textrm{fundamental})=1/2$.
Because we have only bifundamental fields we can write
\begin{equation}\label{acca}
H = 3 N_c - N_f + \sum_i \gamma _i  
\end{equation}
where $N_c $ is the number of colours of the gauge group and $N_f$ the number of flavours.
It is easy to understand that every quiver gauge theory dual to a $Y^{p,q}$ manifold has only four different types of $\beta$ functions given by the four families of gauge factors in figure \ref{beta}.  
\begin{eqnarray}\label{hac}
H_A &=& N^c_A + 2MB_U + \sum_i \gamma ^A_i \nonumber\\
H_B &=& N^c_B + M(B_Y + B_Z ) + \sum_i \gamma ^B_i \nonumber\\ 
H_C &=& M(2B_U + B_Y ) + \sum_i \gamma ^C_i \nonumber\\ 
H_D &=& M(2B_V + B_Y ) + \sum_i \gamma ^D_i  
\end{eqnarray}
where $B_k$ are the baryonic charges of each field as given in the previous section. 
The non-trivial part of the $\beta$ functions is given by the term $\sum _i \gamma _i $ in (\ref{betaf}).
We can compute them using the relations \cite{kleb}
\begin{eqnarray}\label{rel}
2 R_U + R_Y + R_Z &=& 2  \nonumber\\
R_U + R_V + R_Y &=& 2 
\end{eqnarray}
given by the fact that the superpotential has $R$-charge 2, and the following relation between the $R$ charges and the anomalous dimensions
\begin{equation}\label{ranom}
R_i = \frac{2}{3}(1+\gamma _i ) 
\end{equation}  
By doing this we obtain 
\begin{eqnarray}\label{gammai}
\sum _i \gamma _i ^A &=& -N_A + (2B_U - B_Y - B_Z )M + \frac{3}{2} (B_Y R_Y + B_Z R_Z - 2 B_U R_U
)M  \nonumber\\
\sum _i \gamma _i ^B &=& -N_B - (2B_U - B_Y - B_Z )M -\frac{3}{2} (B_Y R_Y + B_Z R_Z - 2 B_U R_U)M  \nonumber\\
\sum _i \gamma _i ^C &=& 2(B_U - B_V )M - 3(B_U R_U - B_V R_V )M  \nonumber\\
\sum _i \gamma _i ^D &=& -2(B_U - B_V )M + 3(B_U R_U - B_V R_V )M  
\end{eqnarray}
Because we will be interested in the leading order in $M/N$ (``near'' the conformal theory), we can use the conformal $R$-charges \cite{kleb}  
\begin{eqnarray}\label{erreconf}
R_Y &=& \frac{1}{3q^2} ( -4p^2 + 2p q + 3 q^2 +  (2p - q )\sqrt{4 p^2 - 3 q^2 })  \nonumber\\
R_Z &=& \frac{1}{3q^2} ( -4p^2 - 2p q + 3 q^2 +  (2p + q )\sqrt{4 p^2 - 3 q^2 }) \nonumber\\
R_U &=& \frac{1}{3q^2} ( 2p( 2p - \sqrt{4 p^2 - 3 q^2 }))
\nonumber\\
R_V &=& \frac{1}{3q} ( 3q- 2p + \sqrt{4 p^2 - 3 q^2 })) 
\end{eqnarray}
to compute the $\beta $ functions:
\begin{eqnarray}\label{betaacca}
H_A &=& (-5p + \sqrt{4p^2 - 3q^2})M  \nonumber\\
H_C &=& (-3(p+q) + \frac{1}{q^2}(4p^3 + 3q^3 - 2pq^2 + (q^2 - 2 p^2 )\sqrt{4p^2 - 3q^2})M  
\end{eqnarray}
and $ H_B = - H_A $, $H_D = - H_C $, where we have also used the explicit expressions for the baryonic charges (see table \ref{bcharges}).
These relations satisfy the following inequalities:
\begin{equation}\label{ineqbet}
H_B > H_D > 0 > H_C > H_A  
\end{equation}  
This means that in the $Y^{p,q}$ gauge theories we always have two $UV$-free families of gauge factors ($(B)$ and $(D)$) and two $IR$-free families ($(A)$ and $(C)$) with opposite $\beta$ functions.

\subsection{Seiberg duality and dimer technology}

\begin{figure}[h]
\centering
\includegraphics[scale=0.7]{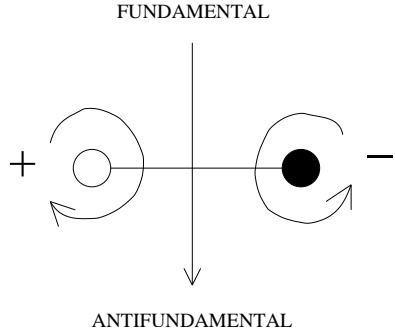}
\caption{The tiling orientation.}\label{orient}
\end{figure}

\begin{figure}[!h]
\centering
\includegraphics[scale=0.4]{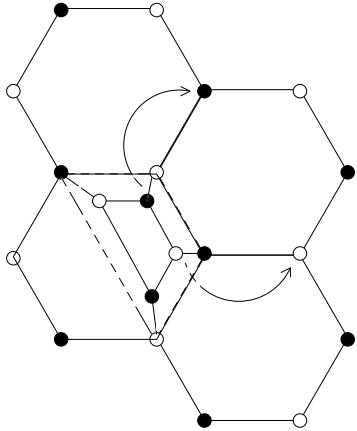}
\caption{Seiberg duality in the brane tiling}\label{sbdual}
\end{figure}

\begin{figure}[!h]
\centering
\includegraphics[scale=0.5]{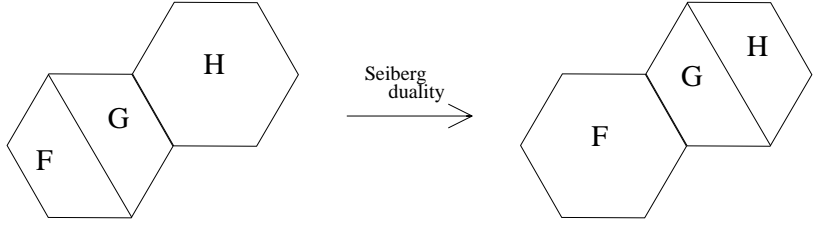}
\caption{Local operation on the dimer}\label{locsab}
\end{figure}

In this section we want to explain how a Seiberg duality can be easily realized in the context of quiver gauge theories using dimer technology \cite{dimers}.
A brane tiling is a bipartite graph in which we can consistently choose an orientation like in figure \ref{orient}. 
This orientation assigns the representation (fundamental or antifundamental) to the chiral superfields in the quiver theory (links).
With this notion in mind we can describe a Seiberg duality at a single gauge group (face) along the renormalization group flow in the following way:
\begin{itemize}
\item{We consider a point in $RG$ such that the dynamics of the other gauge groups is effectively decoupled (all the other gauge couplings can be considered small compared to the one we are dualizing). This reduces the theory to the SQCD-like theory to which Seiberg duality may be applied.} 
\item{Next, we draw inside the cut hexagon a smaller one, we link its vertices, coloured in a consistent way with the orientation, with the vertices of the older one and we erase its edges (fig. \ref{sbdual}; see \cite{dimers}).
This procedure reverses the link orientations, which is the dimer representation of the fact that the dual quarks transform under the conjugate representation to the original ones.
The new links are the mesons $Q_i\tilde{Q}^j = M_i^j $ of the Seiberg duality and they interact with the dual quarks $q_k$ according to the new vertices in the brane tiling giving the usual terms $qM\tilde{q}$.}
\item{Now some of the new fields are massive (terms in the superpotential given by a node with only two arrows), and we have to integrate them out, because we are interested in the low energy theory. 
The integrating out procedure graphically corresponds to erase the massive links and to make the identifications in figure \ref{sbdual}.}
\end{itemize}
 Summarizing, for the types of quiver gauge theories considered in this paper\footnote{To be more precise, this type of operation is the dimer representation of a Seiberg duality that connects two minimal toric phases.}, a Seiberg duality is the local operation on the brane tiling shown in figure \ref{locsab}.
The tools developed in this section are very useful to understand the cascades in the context of very complicated quiver gauge theories as the ones dual to the $Y^{p,q}$ manifolds.

\subsection{$Y^{p,p-1}$ and $Y^{p,1}$ cascades in dimer language }

\begin{figure}[!t]
\centering
\includegraphics[scale=0.47]{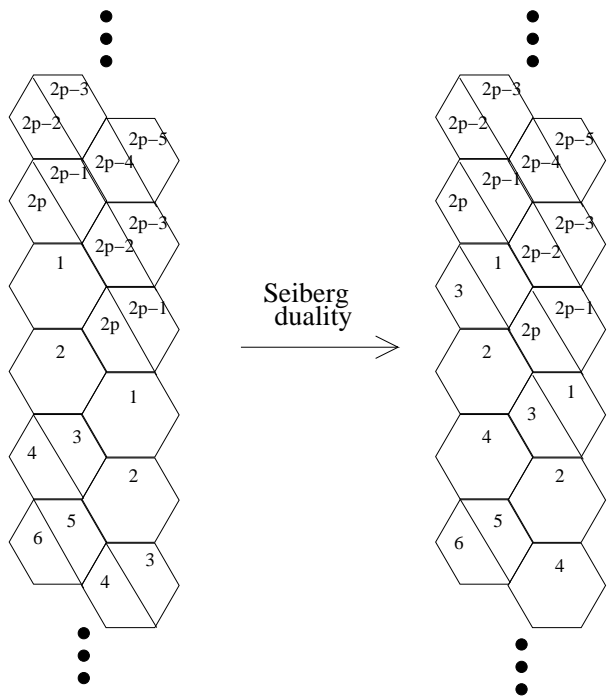}
\caption{The self-similar structure of the $Y^{p,1}$ quiver gauge theories under Seiberg duality.}\label{yp1sd}
\end{figure}

In \cite{kleb} the cascades for the dual gauge theories to $Y^{p,p-1}$ and $Y^{p,1}$ manifolds are explained.
Using the dimer technology we can show in a simpler way the self-similar structure of the gauge theory under duality, which allows the cascade to occur.
The case of $Y^{p,p-1}$ is quite general: the theories have only one cut hexagon and $n=2(p-1)$ regular hexagons. Because we have only one gauge factor with the largest $UV$-free $\beta$ function (type $(B)$) we can choose the most generic initial conditions in which this factor (that is also the one with the highest rank) has the strongest gauge coupling. In the present situation this gauge coupling will run to infinity first and we have to apply the duality transformation to this gauge group factor if we want to continue the $RG$ flow after this point.      
\begin{figure}[!h]
\centering
\includegraphics[scale=0.45]{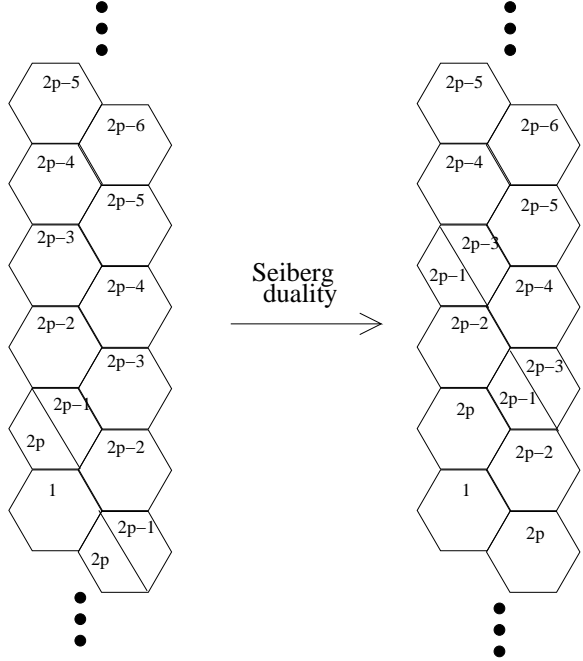}
\caption{The self-similar structure of the $Y^{p,p-1}$ quiver gauge theories under Seiberg duality}\label{ypp1sd}
\end{figure}
In figure \ref{ypp1sd} the self-similar structure of the theory is evident. In fact we obtain the original theory if we relabel the gauge factors in the following way:
\bea
\label{relabypp1}
2p \rightarrow 2 \quad 2p-1 \rightarrow 2p  \quad 2p-2 \rightarrow 1 \quad 2p-3 
\rightarrow 2p-1 \nonumber \\ \quad  2p-4 \rightarrow 2p-2 \quad  2p-5 \rightarrow 2p-3 \quad \ldots
\eea
The net effect on the theory is the shift $N \rightarrow N - M $. At the next step we will need to dualize the ``new'' $2p-1$ factor, and so on.
The case of $Y^{p,1}$ is less generic than the previous one because these theories have $n=2$ regular hexagons and $ m=p-1 $ cut hexagons. This means that a generic theory will have $p-1$ type $(B)$ beta functions and there may be different natural choices to realize the self-similar structure of the theory under $RG$ flow. The easiest one is to choose the initial conditions such that the gauge group factor of highest rank flows to infinite coupling first, then after applying a duality transformation to this group and relabeling the group factor, we find exactly the same theory with $N\rightarrow N-M $.  
The procedure is shown in figure \ref{yp1sd}.
In this case the relabeling is:
\begin{equation}\label{relabyp1}
3 \rightarrow 2p \quad 1 \rightarrow 2p-1  \quad 2p \rightarrow 2p-2 \quad 2p-1 \rightarrow 2p-3 \quad 2 \rightarrow 1 \quad  4 \rightarrow 2 \quad ... \nonumber\\
\end{equation}
At the next step we will need to dualize the ``new'' $3$ factor and so on.

\begin{figure}[!h]
\centering
\includegraphics[scale=0.55]{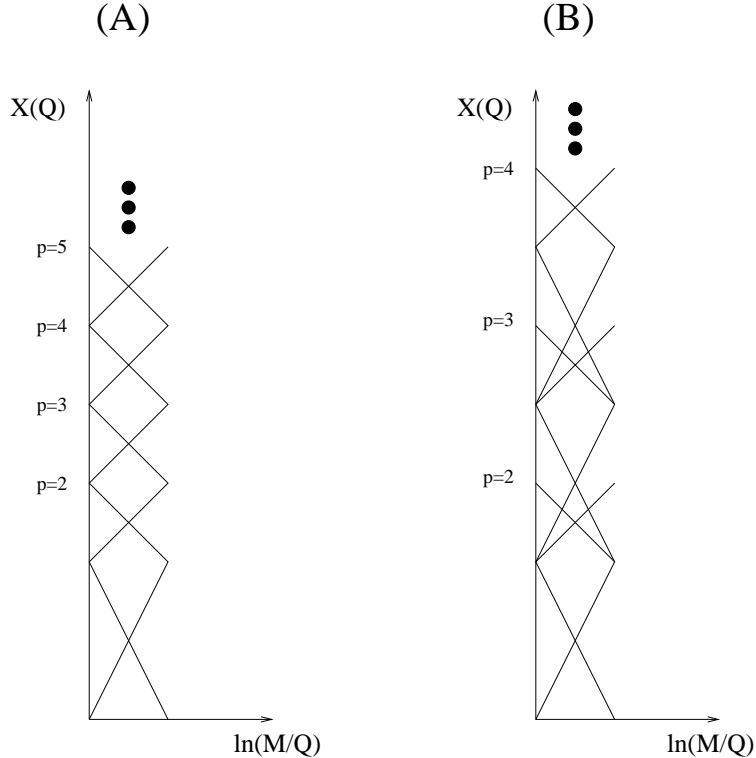}
\caption{The general structure of the $RG$ flow for $Y^{p,p-1}$ $(A)$ and for $Y^{p,1}$ $(B)$}\label{Ypp-1Yp1}
\end{figure}

Until now we have shown possible self-similar patterns of gauge dualities, and we have justified them giving some arguments about the ``naturalness'' of the relative sizes of the gauge coupling constants. 
To show that such cascades really occur and that they are consistent with the renormalization group flow we must explicitly write down the possible pattern of initial data for the gauge coupling constants\footnote{We will always work in the limit $N \gg M$ that allows us to use the conformal anomalous dimensions in the computation of the $\beta$ function.}.
In what follows we will use a graphical method: we will plot the $RG$ flow for the coupling constants as a function of the logarithm of the energy $\ln ( M / Q )$:
\begin{equation}\label{costin}
X(Q) = X_M - H \ln ( M / Q ) 
\end{equation}  
where $M$ is the $UV$ starting point of the $RG$ flow, $Q$ is the energy scale along the flow, and $X(Q)= \frac{8 \pi }{g^2 ( Q )}$, $ X_M = \frac{8 \pi }{g^2 ( M )}$. \\
Let us start with the very easy example of the gauge theory dual to the $Y^{2,1}$ manifold which is common to both the categories.
\begin{figure}[!b]
\centering
\includegraphics[scale=0.7]{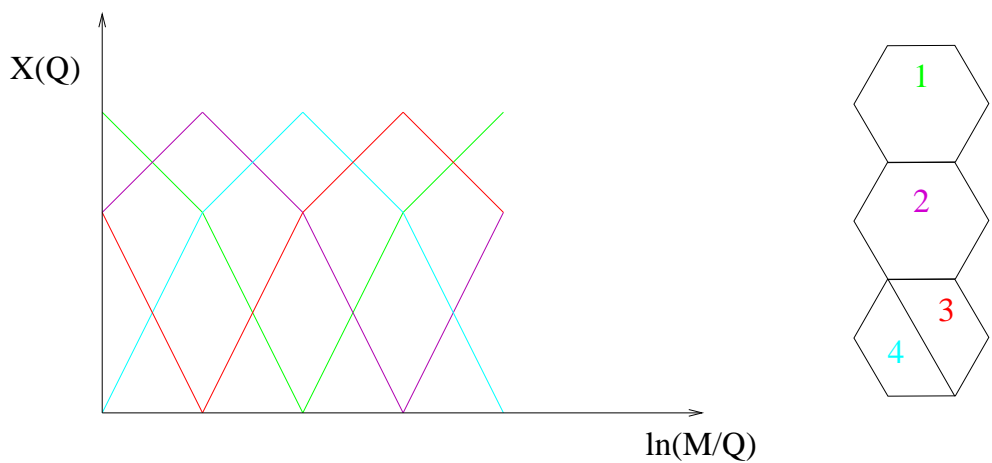}
\caption{The complete period for the $RG$ flow of $Y^{2,1}$ gauge coupling constants.}\label{y21cos}
\end{figure}
In figure \ref{y21cos} it is easy to see the presence of two periods: the first one is one step long and relabeling the gauge groups we find this period repeated at every Seiberg duality, while the second one is four steps long: after four steps the theory returns to itself with the same gauge groups exactly in the same positions. \\
Now we want to explain the general structure of the $RG$ flow for $Y^{p,p-1}$ gauge theories, starting for pedagogical purposes with the example of $Y^{3,2}$.
\begin{figure}[!b]
\centering
\includegraphics[scale=0.65]{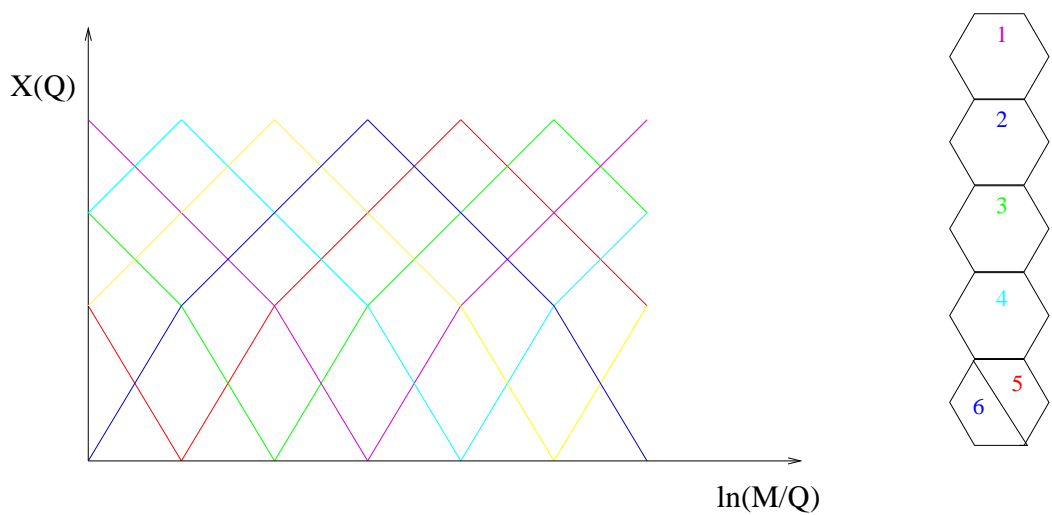}
\caption{The complete period for the $RG$ flow of $Y^{3,2}$ gauge coupling constants}\label{Y32cos}
\end{figure}
Also in this case we find two types of periods, a one step long and another one which is as long as the number of gauge groups of the theory.
This is a general feature.
Analysing the $Y^{p,p-1}$ case in general we find:
\begin{itemize}
\item{$\beta _B / \beta _D \sim 2$ }
\item{$p \rightarrow p+1 $ $\Rightarrow $ $n \rightarrow n+2$ }
\item{every gauge group stays in the type $(B)$ position for just one step}
\item{every gauge group stays in the type $(A)$ position for just one step}
\item{every gauge group stays in the type $(C)$ position for $n/2$ steps}
\item{every gauge group stays in the type $(D)$ position for $n/2$ steps}
\item{the theory returns to itself after $n+2 = 2p = \# \hbox{gauge groups}$ steps }
\end{itemize}
This analysis gives us the general picture for the $RG$ flow of $Y^{p,p-1}$ theories (fig. \ref{Ypp-1Yp1}).

\begin{figure}[!b]
\centering
\includegraphics[scale=0.6]{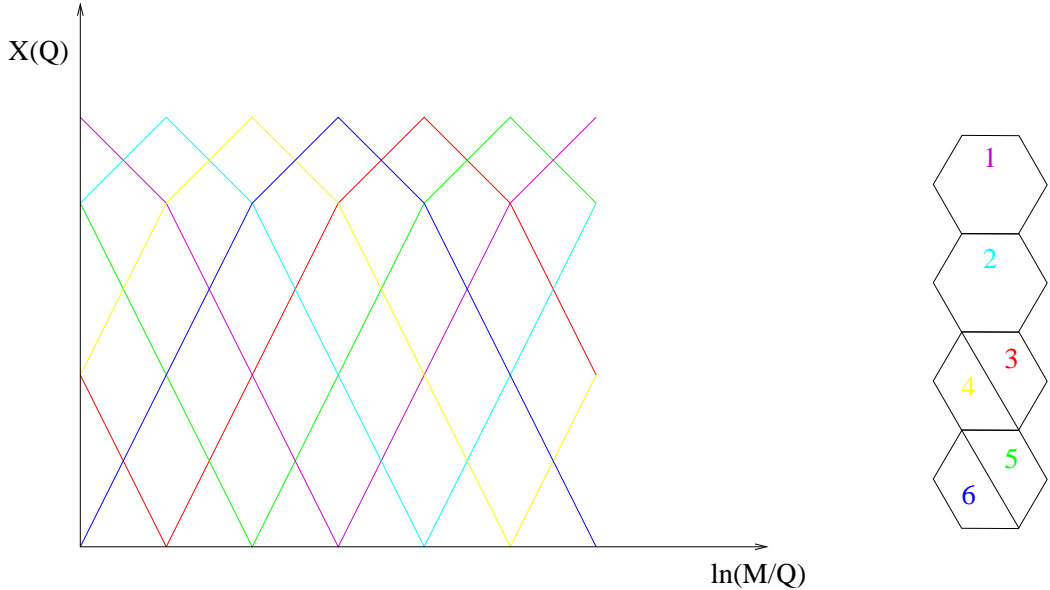}
\caption{The complete period for the $RG$ flow of $Y^{3,1}$ gauge coupling constants}\label{Y31cos}
\end{figure}

To explain the general features of the $RG$ flow of $Y^{p,1}$ gauge theories it is useful to start again with an example: $Y^{3,1}$ (fig. \ref{Y31cos}).
As usual we find two periods: the short one and the long one.
The results for the generic $Y^{p,1}$ theory are the following:
\begin{itemize}
\item{$\beta _B / \beta _D \sim 2$ }
\item{$p \rightarrow p+1 $ $\Rightarrow $ $m \rightarrow m+1$ }
\item{every gauge group stays in the type $(B)$ position for $m$ steps}
\item{every gauge group stays in the type $(A)$ position for $m$ steps}
\item{every gauge group stays in the type $(C)$ position for just one step}
\item{every gauge group stays in the type $(D)$ position for just one step}
\item{the theory returns to itself after $2m+2 = 2p = \# \hbox{gauge groups}$ steps}
\end{itemize}  
Using this analysis we arrive at the general picture for the $RG$ of $Y^{p,1}$ gauge theories (figure \ref{Ypp-1Yp1}).

\subsection{Remarks on the cascades for the generic $Y^{p,q}$ manifold}

\begin{figure}[!b]
\centering
\includegraphics[scale=0.62]{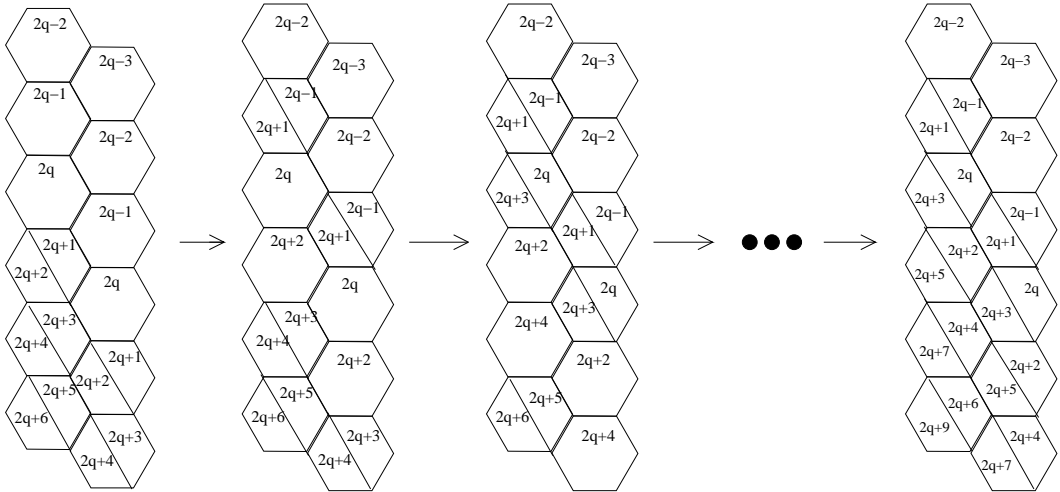}
\caption{The first self-similar pattern of dualities for $Y^{p,q}$.}\label{ypqsdc1}
\end{figure}

\begin{figure}[!b]
\centering
\includegraphics[scale=0.55]{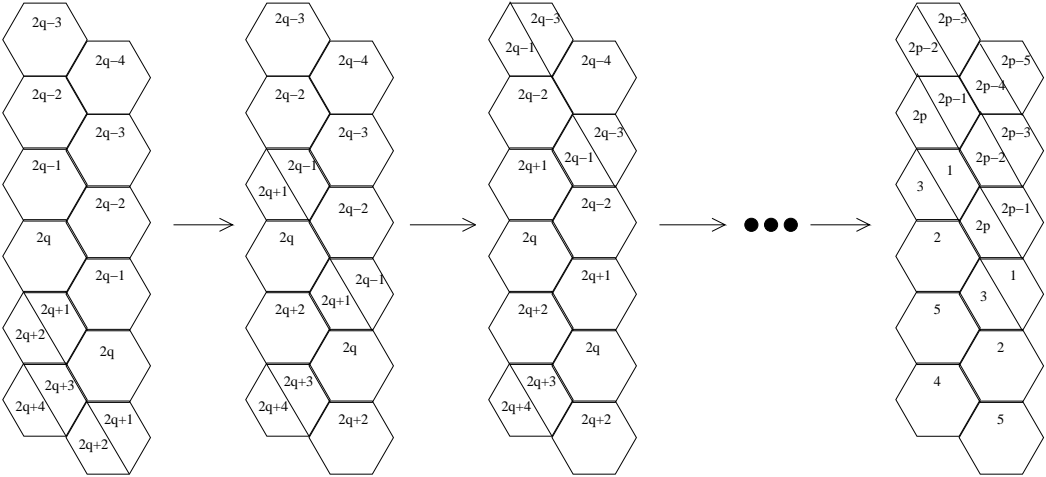}
\caption{The second self-similar pattern of dualities for $Y^{p,q}$. }\label{ypqsdc2}
\end{figure}

Now that we have understood the cascades for the two subfamilies of gauge theories dual respectively to $Y^{p,p-1}$ and  $Y^{p,1}$ manifolds we can try to guess the cascades for the entire $Y^{p,q}$ family. 
The main support for the idea of cascading $RG$ for the gauge theories associated to the $Y^{p,q}$ singularities comes from the supergravity dual description \cite{kleb}.
In fact it was found that the five form $F_5$, whose integral over the $Y^{p,q}$ manifolds gives the number $N$ of regular branes in the conformal case, acquires a radial dependence in presence of additional fractional branes that is a monotonically increasing function of the radius $r$ of the cone.
Considering the relation between the radial coordinate $r$ of the geometry and the energy $E$ of the dual gauge theory this means that going toward the $IR$ (small values of $r$) the number of effective regular $D3$-branes decreases and this situation can be realized naturally in the gauge field theories as a sequence of Seiberg dualities. Thanks to the presence of quartic terms in the superpotential the renormalization group flow can be self-similar \cite{strassler}.\\  
As we have seen the $Y^{p,p-1}$ and the $Y^{p,1}$ cascades are examples in which we can immediately see the self-similar structure of the theory under duality after a single step of the cascade.
Obviously this is not the general case. To show the plausibility of the existence of a cascade we need to discover a pattern of Seiberg dualities that allows us to reconstruct the original theory, doing a simple relabeling of gauge factors, after a finite numbers of steps.
In the literature (see \cite{hananyuranga2} and references therein) many examples are known in which the ``period'' of the cascades is greater than one and the shift is not as simple as ``$N\rightarrow N-M$''. Some examples that have these properties are the dual gauge theories of the complex cone over the Hirzebruch surface $F_0$, the complex cone over the second del Pezzo surface $dP_2$, the suspended pinch point singularity $SPP$ and many others.
Looking at the brane tiling representing the gauge theory dual to a generic $Y^{p,q}$ manifold it seems reasonable to propose that the situation for the general case could be similar to the ones we have just explained, namely the self-similar structure is realized after a number of Seiberg dualities bigger than one.
We can in fact easily find two different simple patterns of dualities that show explicity their self-similar structure.
The first one is explained in figure \ref{ypqsdc1}: after $m$ steps, relabeling the gauge group factors in the following way:
\begin{equation}\label{relabypq1}
2q-2 \rightarrow 2q \quad 2q-1 \rightarrow 2q+1  \quad 2q+1 \rightarrow 2q+2 \quad 2q \rightarrow 2q+3 \quad 2q+3 \rightarrow 2q+4 \quad ... \nonumber\\
\end{equation}
we obtain the original tiling with an overall shift in the ranks of the gauge groups factors equal to minus the number of steps:
\begin{equation}\label{rescypq1}
N \rightarrow N - (p-q)M
\end{equation}
The second one is explained in figure \ref{ypqsdc2}.
After $n/2$ steps relabeling the gauge group factors in the following way:
\begin{equation}\label{relabypq2}
5 \rightarrow 2 \quad 2 \rightarrow 1  \quad 3 \rightarrow 2p \quad 1 \rightarrow 2p-1 \quad 2p \rightarrow 2p-2 \quad ... \nonumber\\
\end{equation}
again we obtain the original tiling with an overall shift in the ranks of the gauge groups factors now equal to:
\begin{equation}
N \rightarrow N -qM
\end{equation}
These two types of patterns are clearly degenerate in the cases of $Y^{p,p-1}$ and $Y^{p,1}$ and are realized in a single step.\\
To claim that these self-similar patterns reproduce a duality cascade it remains to show that it is possible to choose initial couplings such that the proposed dualities take place along the $RG$ flow. 
To attack the problem of finding a disposition of initial gauge couplings, we used the quiver gauge theory dual to $Y^{5,2}$ as a toy model for the complete family (as a tiling of $T^2$ it is general enough to give us some feeling for the general case). \\
Up to now we have shown only the most obvious sequences of Seiberg dualities that make the theories return to themselves after a finite number of steps. There are many other possible periodic patterns in which we consider also simultaneous dualization of more than one gauge group factor.
The examples of cascades dual to supergravity solutions present in literature have Seiberg dualities occurring at constant intervals in $\ln \mu $, but the supergravity solution tells us only that the decreasing of the number of effective regular branes is linear in $\ln \mu $ \cite{hanany}. This implies that in the most general cases there exist periodic flows with a constant interval in which the theory returns to itself without any permutation (large step), while inside this step the Seiberg dualities can be arbitrarily distributed in the energy ranges. 
We considered some of  these general possibilities for a class of self-similar patterns in the specific case of $Y^{5,2}$, but we didn't succeed. In fact, even allowing dualizations of more than one gauge group simultaneously and varying the steps in energy of the dualities inside a long period we could not find any possible disposition of the initial values that realize a periodic flow of the coupling constants. The analysis is clearly not complete, but it seems that contrary to the known cases the $RG$ of the theory does not follow the symmetries of the dimer.
The possible solution to this problem may be related to the fact that until now we have considered only minimal toric phases of the theory, while in more general situations the cascade needs Seiberg dualities between generic toric phases (for example the cascade of the gauge theory dual to the complex cone over $F_0$).
A toric phase of a quiver gauge theory is a phase in which in the conformal regime all the gauge group factors are $SU(N)$ with the same rank. Flowing in the $IR$ in the non-conformal case the only information we have from the supergravity solution is that the number of effective $D3$ regular branes must constantly decrease. This can be realized in general flowing through all the toric phases of the dual gauge theory. In fact in a generic point of the $RG$ flow any factor of the gauge group has the form $SU(N + \alpha M)$ with $\alpha $ an integer number. Seiberg dualities inside the set of toric phases do not change the number $N$ of initial $D3$-branes affecting only the term proportional to $M$, while Seiberg dualities that link toric phases with non-toric ones make the term proportional to $N$ increase. For this reason along the cascading $RG$ flow we expect only toric phases, but in general also non-minimal ones.
Non-toric phases have more chiral fields than the minimal ones and in \cite{kaza} it is explained how to construct all the connected toric phases for the complete class of $Y^{p,q}$ quiver gauge theories.
It seems reasonable that to realize the $RG$ flow dual to the supergravity solution for the entire class of $Y^{p,q}$ manifolds we need to pass through non-minimal toric phases\footnote{We want to thank Amihay Hanany for useful comments about this point.}.     
This last point clearly needs more study.

\section{Dynamical Supersymmetry Breaking and runaway behaviour: some comments}
\label{DSBrun}
Our next task is the study of the $IR$ behaviour for the entire class of gauge theories dual to $Y^{p,q}$ manifolds at the end of the cascade with $p>q>0$. 
As explained in section \ref{GeneralYpq} the algebraic structure of these spaces suggests that the non-perturbative dynamics breaks supersymmetry. 
The point is that until now there are no tools in toric geometry or in dimer technology to understand if a stable non-supersymmetric vacuum exists ($DSB$) or supersymmetry is in some sense restored at the infinity in the space of vevs of the fields (runaway behavior). 
Traditionally, one of the requirements for theories with $DSB$ is that they do not possess classically flat directions. This requirement was due to the observation that when classically flat directions are lifted by non-perturbative dynamics one usually finds a runaway behavior (see \cite{shir} and references therein). This is because supersymmetry can be broken due to a variety of effects, but in general it is the consequence of the interplay between, on the one hand, a tree-level superpotential, which gives rise to a non-zero potential everywhere except at the origin in the space of fields, and, on the other hand, non-perturbative effects, which generate a potential that is non-zero at the origin\footnote{This is the traditional situation, but there are also cases in which quantum effects can stabilize the classically flat directions yielding $DSB$ also in presence of flat moduli (see \cite{shir,shir2} and references therein ).}.
In \cite{hananyuranga} it is proven that all the gauge theories dual to $Y^{p,q}$ with $p>q>0$ have a gauge group factor with $N_f < N_c$ which hence generates an $ADS$ superpotential. This suggests that supersymmetry is broken. In the following we will give a complete classification of the pattern of the $ADS$ superpotentials, explaining that generically the number of gauge group factors that develop a non-perturbative $ADS$ superpotential is more than one.\\
As an introduction to the problem we want to explain the general feeling about these theories. 
To understand the IR behavior we will consider the theories at the end of the cascades, where there are only fractional branes. At low energy the $U(1)$ factors are massive and hence decoupled leaving only $SU(N)$ factors. In this situation we have two types of vevs: mesonic and baryonic. The former are given by the generators of the mesonic chiral ring and are in a one to one correspondence with the embedding coordinates of the singular algebraic intersection. Actually there exists a precise map betweeen the set of generators of the dual lattice cone (embedding coordinates) and the generators of the mesonic chiral ring (mesonic chiral primaries) \cite{berenstein,pin,hanher,sergio}. This tells us that we will not have mesonic flat directions, because they are related to the coordinates of the cone and fractional branes must sit at its tip (the origin of the embedding space) imposing the vanishing of all the vevs of the mesonic chiral primaries. In this way we have at our disposal as runaway directions only the baryonic ones; indeed, we will see that it is always possible to construct a baryonic classically flat direction that is lifted to runaway by non-perturbative effects. This is what allows us to believe that the complete dynamics is runaway\footnote{The interesting fact is that if we study the theory with $U(N)$ factors (this corresponds in some sense to consider the correspondence in the string regime, before the low energy limit) we can turn on $FI$ terms for the $U(1)$. In the conformal regime their number is $\# \textrm{gauge-groups} -1$ and they are related to the possible ``deformations'' of the supergravity solution on the geometry side and to non-baryonic operators on the field theory side \cite{sergio2}. In this framework, runaway seems to be given by the fact that closed string fields at the singularity are dynamical and couple as $FI$ terms in the gauge sector. These fields potentially stabilize the runaway lifting the classically flat directions thanks to $D$-terms conditions, but the complete dynamical process suggests us that at the end we have runaway. In fact the $FI$ fields in string compactifications have no potential and the system relaxes pushing these fields to infinity restoring the runaway directions \cite{hananyuranga} that in the low energy field theories language are baryonic directions. It would be interesting to understand if there is any relation between these $FI$ terms and baryonic flat directions in general in the non-conformal regime.}.
Clearly the complete dynamics of the theory is very complicated and it might happen that the K\"ahler potential stabilizes the otherwise runaway directions.\\
Before starting our analysis of the $Y^{p,q}$ family, we want to make some comments about the cases discussed in literature of quiver gauge theories that are dual to geometries that allow $SB$ branes. In these situations an $ADS$ superpotential is generated that lifts the otherwise classical baryonic flat direction and gives a runaway behavior to the theory.  
We want to emphasize the simple structure of the superpotential in these cases, and why for this reason it is worthwhile to discuss more complicated cases in which the structure of the superpotential is more general and the $IR$ dynamics is richer, developing for example more than one $ADS$ superpotential term.\\
Let us discuss some examples:
in this short review we will be very schematic and we refer to the original literature (see \cite{hananyuranga2} and references therein) for the discussion of the structure of these theories. We will denote with $V$ the possible elementary fields associated to runaway directions, $M$ the composite fields that are part of the meson field related to the node which develops the $ADS$ superpotential, and $\mathcal{M}$ the meson constructed with them.
We will not care about the numerical factors since the only thing we are interested in is the form of the complete effective superpotential.
We would only like to point out the simplicity of the interaction structure of the fields that gives the runaway behavior for the known quiver theory which admit $SB$ branes.  
\subsubsection{$Y^{2,1}$}
 
The effective superpotential is given by:
\begin{equation}\label{sp21}
W_{\hbox{eff}}=M^1 V^1 + M^2 V^2 + \frac{\Lambda}{\det \mathcal{M}}
\end{equation}
The $F$-term equations for the $V^i$ fields put the vev of the $M^i$ fields to zero and these push the vev of the $V^i$ fields to infinity.
It easy to write a gauge invariant operator with these two fields:
\begin{equation}\label{ginv21}
\epsilon_{i_1 ... i_{2M}}\epsilon_{j_1 ... j_M}\epsilon_{k_1 ... k_M}V^1_{j_1 i_1}...V^1_{j_M i_M}V^2_{k_1 i_{M+1}}...V^2_{k_M i_{2M}}
\end{equation}
This in fact parametrizes the runaway direction.

\subsubsection{$dP_2$}

The quiver gauge theory dual to $dP_2$ singularity with fractional branes given by the vector $M(1,0,1,0,2)$ has the effective superpotential:
\begin{equation}\label{sp2}
W_{\hbox{eff}}=MV + \frac{\Lambda}{\det M}
\end{equation}
The $F$-term equations put to zero the vev of the meson and push to infinity the elementery field $V$.
We can easily construct a gauge invariant baryonic operator
\begin{equation}\label{ginvdp2}
\epsilon_{i_1 ... i_M}\epsilon_{j_1 ... j_M}V_{i_1 j_1}...V_{i_M j_M}
\end{equation}
This is the runaway direction.

\subsubsection{$dP_3$}

This singularity admits three different types of fractional branes. The $SB$ ones are given by the vector $K(1,0,1,0,1,0) + P(1,0,0,1,0,0)$ with $P \ll M$ we obtain again an effective superpotential of the form:
\begin{equation}\label{sp3}
W_{\hbox{eff}}=MV + \frac{\Lambda}{\det M}
\end{equation}
 As usual the $F$-term equations push the value of $V$ to infinity and we can construct a baryonic gauge invariant as in the previous subsection.

\subsubsection{The general case}

In general the situation is more intricate.
We can have an effective superpotential of the form:
\begin{equation}\label{gen}
W_{\hbox{eff}}=M^1V^1 + M^2V^2 + \frac{\Lambda}{\det \mathcal{M}} + V^1 X^1 X^2 + V^2 X^3 X^4 + ...
\end{equation}
where $X^k$ are other fields in the theory.
This is the case with only one $ADS$ non-perturbative superpotential, but as we will see in the general case we will have more of them.
In the following sections we will try to give arguments for the runaway behaviour of these more complicate theories by studying the class of $Y^{p,q}$ singularities and some more general examples concerning the $L^{a,b,c}$ singularities.

\section{The IR regime of non-conformal $Y^{p,q}$ theories: general features}\label{Ypqdimer}

As explained in the previous sections, most of the properties for the whole $Y^{p,q}$ family can be easily read off the associated dimer 
graph. The reason why a general analysis can be carried out is related to the peculiarly simple 
properties of the tiling for this class of manifolds, allowing us to make general statements about the gauge structure at the end of the cascade and the way non-perturbative runaway terms may be generated in the low energy effective superpotential.

\subsection{The assignment of the ranks in the non-conformal regime}

\begin{figure}[!b]
\centering
\includegraphics[scale=0.8]{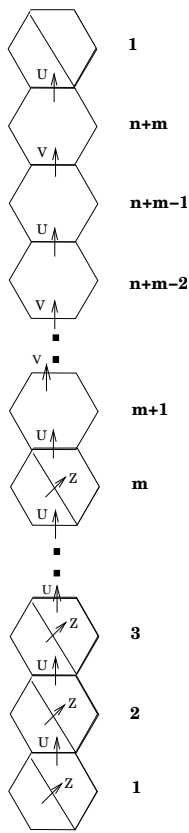}
\caption{Assignment of $V$, $U$ and $Z$ fields inside the tiling for $Y^{p,q}$.}
\label{ypqfields}
\end{figure}

In section \ref{GeneralYpq} we outlined some important features of the dimer graph for the $Y^{p,q}$ family. First and most important, we have $k=1$ periodicity in the transverse direction; secondarily, as shown in \cite{zaffa}, we are allowed to put all the $m$ divided hexagons consecutively. Following \cite{zaffa}, the general pattern for assigning $U$ or $V$ fields to vertical arrows is then the one of figure \ref{ypqfields}.
For all these reasons, and especially due to the $k=1$ periodicity, the number of colors for each 
factor group of the dual gauge theory is assigned in a very simple fashion following the recipe explained in section \ref{GeneralYpq}. Let $N$ be the rank 
of the lowest-lying face of the whole tiling, which, by the above discussion, is the left half of 
a divided hexagon. Then denote the quadrilaterals of the divided hexagons by an index $i=1 \ldots 2m$ 
starting from the bottom one, and label the regular ones as well with an another ascending index 
$j=1 \ldots n$ where $j=1$ corresponds to the face just above the tower of cut hexagons. According to table \ref{bcharges} and making reference to figure \ref{ypqfields}, it is easily seen that the number of colors are assigned as follows:
\begin{equation}
\label{eqdisacc1}
\hbox{Divided hexagons:} \qquad N^{div}_c(i) = 
\left\{
\begin{array}{lc}
N + [(i/2)q + p]M & \quad \hbox{$i$ even}\\
N +  [(i-1)q/2]M & \quad \hbox{$i$ odd} \\
\end{array}\right.
\end{equation}
\begin{equation}
\label{eqdisacc2}
\hbox{Regular hexagons:} \quad N^{reg}_c(j) = 
\left\{
\begin{array}{lc}
N + [(p-q)(q+1-j/2) +q]M  & \quad \hbox{$j$ even} \\
N + (p-q) [q + (1-j)/2 ]M & \quad \hbox{$j$ odd} \\
\end{array}\right.
\end{equation}

\subsection{Effective low energy structure via dimer analysis: gauge decoupling and ADS superpotentials}

\begin{figure}[!!h]
\centering
\includegraphics[scale=0.75]{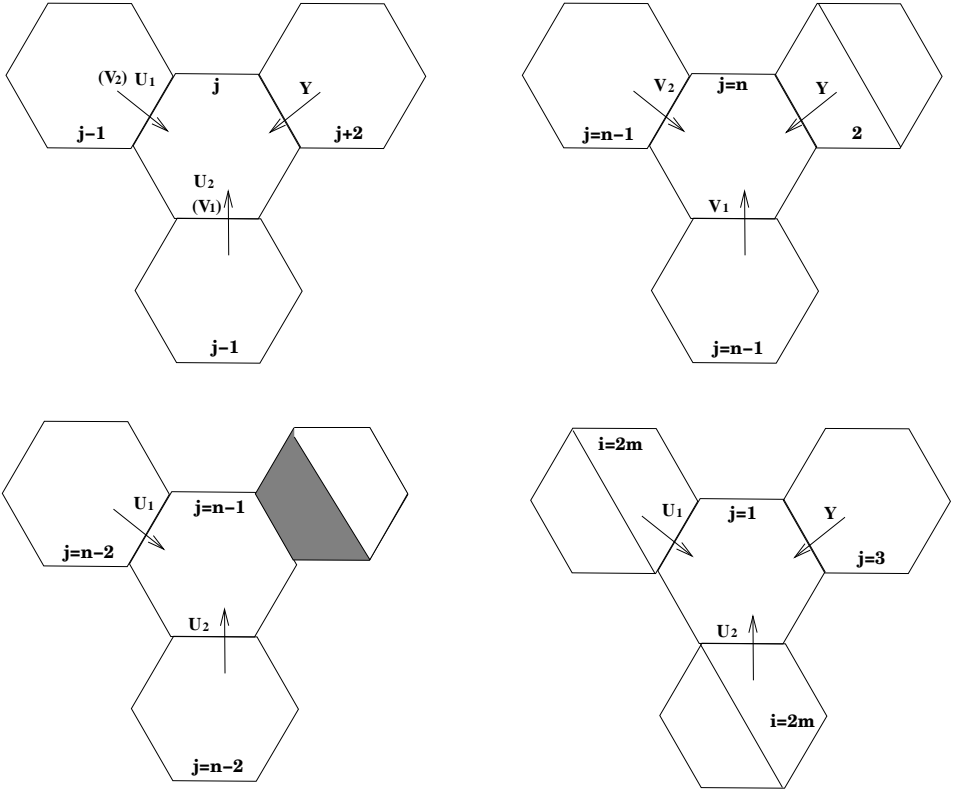}
\caption{The ``bulk'' (top-left) and the three ``boundary'' regular hexagons ($m=p-q$).}
\label{bulkboundreg}
\end{figure}

Suppose now we are at the end of a cascade of Seiberg dualities, thus setting\footnote{In the last step of the cascade we should be more careful and take into account the quantum dynamics of the theory, but it seems that this whole process is equivalent to perform naive Seiberg duality.} $N=0$. Our task in this section is to classify the gauge group factors that decouple in the $IR$ ($N =0$) and the ones that develop an $ADS$ superpotential ($0<N_f < N_c$). As a first step in our analysis we notice from (\ref{eqdisacc1}) and (\ref{eqdisacc2}) that the
number of colors of the gauge group associated to the bottom-left half of the first cut hexagon is strictly smaller than $N^{div}_c(i)$ and $N^{reg}_c(j)$ for all $j$'s and $i >1$. Thus we see that for this class of theories only the lowest gauge factor in the dimer graph decouples at the end of the cascade. \\
Moreover, it is straightforward to figure out the conditions that ensure that one or more of the $SU(N_i)$ factors develop a non-perturbative superpotential of the Affleck-Dine-Seiberg type. Let us consider the case of regular hexagons first. To simplify the notation we provisionally put $M=1$ in the following formulas as it will
have no influence on relations such as $N_f \lesseqgtr N_c$. To carry out the analysis we distinguish between ``bulk'' and ``boundary'' hexagons (figure \ref{bulkboundreg}), the former being the ones whose incoming 
arrows come only from other regular hexagons (i.e., their fundamental matter fields are charged
under the antifundamental of uncut factors only, so they belong to the portion of the tiling
consisting only of regular hexagons not bordering any divided one), while the latter may have
matter with ``flavor'' indices belonging to the antifundamental of some cut hexagon.
For ``bulk'' hexagons, we further distinguish between even and odd ones according to $j$ being
even or odd respectively. In the odd case, we have a $U^\alpha$ doublet coming from the
$(j-1)^{th}$ regular hexagon and a $Y$ field from the $(j+2)^{th}$. We can then easily compute the
number of flavors exploiting (\ref{eqdisacc2})
\begin{equation}
N^{reg}_f(j)=2N^{reg}_c(j-1)+N^{reg}_c(j+2)= 3N^{reg}_c(j)+p+q \qquad \hbox{($j$ odd)}
\end{equation}
so they never lead to the appearance of an ADS term in the quantum superpotential. The same
computation applied to the even case gives
\begin{equation}
N^{reg}_f(j)=3N^{reg}_c(j)-p-q \quad \hbox{($j$ even)}
\end{equation}
and using (\ref{eqdisacc2}) we are eventually led to
\begin{equation}
N^{reg}_f(j)- N^{reg}_c(j)=2(p-q)(q-j/2+1)+q-p \geq 3(p-q) > 0 \quad \hbox{($j$ even)}
\end{equation}
Again no ADS superpotential is generated. As far as the  ``boundary'' hexagons are concerned we 
have
\begin{eqnarray}
N^{reg}_f(2q) &=& N^{div}_c(3)+2N^{reg}_f(2q-1) = q+2(p-q) =\\
&=& 2p-q>p=N^{reg}_c(2q)  \nonumber \\
N^{reg}_f(2q-1) &=& 2N^{reg}_c(2q-2) = 2(2p-q)>p-q=N^{reg}_c(2q-1) \nonumber \\
N^{reg}_f(1) &>& 2N^{div}_c(p-q) = 2(N^{reg}_c(1)+p) > N^{reg}_c(1)
\end{eqnarray}
and we conclude that in no case regular hexagons give rise to ADS-type non-perturbative
contribution to the superpotential. \\
Let us then consider the case of divided hexagons. First of all it is rapidly checked that the
top-right half of the lowest hexagon, whose bottom-left half has undergone decoupling, does indeed
generate an ADS superpotential
\begin{equation}
N^{div}_c(2) = p+q, \quad N^{div}_f(2)=2q < N^{div}_c(2)
\end{equation}
As far as the other cut hexagons are concerned, we only focus on quadrilaterals of even index $i$ since
they are the only ones that have a chance to have $N_f < N_c$: the odd ones have just two $U$ fields entering, which have baryonic charge equal to $-p$, and hence $N_f > N_c$.. The distinction between ``bulk''
and ``boundary'' is now immaterial since in both cases we have, for the $i^{th}$ cut hexagon, an
incoming $Y$ field carrying a  $N^{div}_c(i) -(p-q)$ flavor index and a $Z$ field with
$N^{div}_c(i) - p-q$ flavor symmetry. Therefore, for $i=2,\ldots,m$ we have that
$$N^{div}_c(2i)=p+iq $$
$$N^{div}_f(2i)=2N^{div}_c(2i) -2p$$
so that the gauge factors which develop an ADS-type superpotential are given by
\begin{eqnarray}
N^{div}_f(2i)-N^{div}_c(2i) &=& N^{div}_c(2i)-2p = \nonumber \\
&=& iq-p < 0
\end{eqnarray}
that is to say
\begin{equation}
i < \frac{p}{q} \qquad i=2,\ldots,m
\label{eqADS}
\end{equation}
Summing up, we are led to the following conclusions for the entire $Y^{p,q}$ family with $p>q>0$:
\begin{enumerate}
\item only one of the gauge factors, corresponding to the bottom-left half of the lowest-lying hexagon, decouples at the end of the cascade;
\item no uncut hexagon gives rise to ADS-like terms in the quantum superpotential;
\item gauge factors related to divided hexagons may develop non-perturbative contributions to $W_{eff}$ according to (\ref{eqADS})
\end{enumerate}

\section{Analysis of the IR behavior}
\label{YpqIR}
We now proceed to the study of the vacuum structure of the low energy theory for generic $p$ and $q$ with $p>q>0$. We will primarily concentrate on the analysis of the $F$-terms for the effective superpotential $W_{eff}$. We will not try a complete analysis of the vacuum structure of these field theories. Our task is to understand whether the $SB$ fractional branes allowed by the geometry are of $DSB$ or runaway type. With this aim in mind we are going to show that it is always possible to construct a gauge invariant classically flat baryonic direction which is turned to runaway by the $ADS$ non-perturbative contribution to the superpotential.

\subsection{$p-q < \frac{p}{q}$}
\label{mes}
\begin{figure}[!b]
\begin{minipage}[t]{0.49\linewidth}
\centering
\vspace{0pt} 
\includegraphics[scale=0.65]{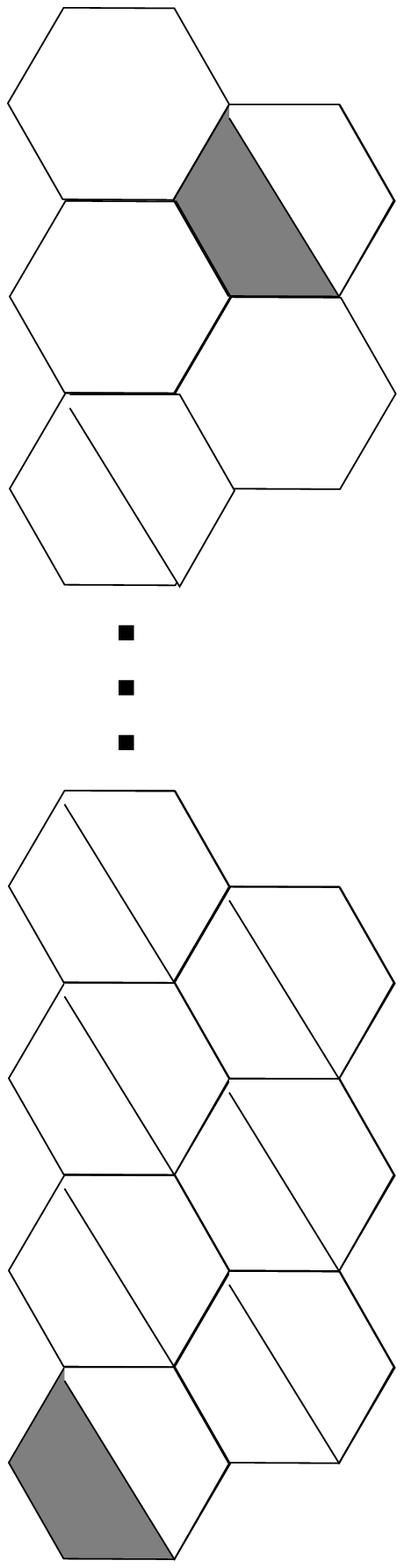}
\vspace{0pt} 
\caption{The dimer for $Y^{p,1}$}
\label{yp1fig}
\end{minipage} 
\begin{minipage}[t]{0.49\linewidth}
\centering
\vspace{0pt} 
\includegraphics[scale=0.65]{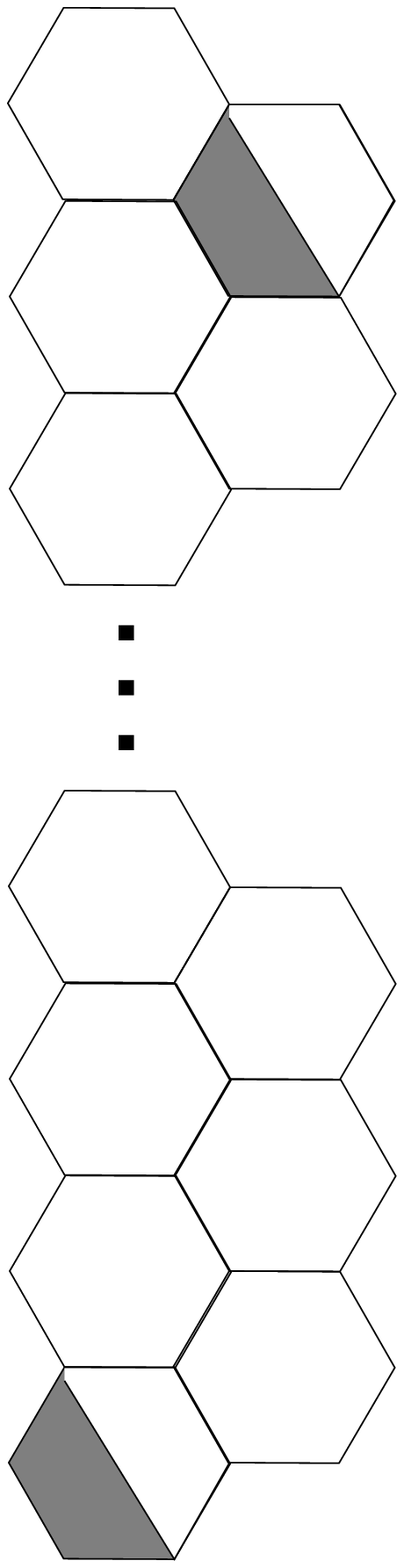}
\vspace{0pt} 
\caption{The dimer for $Y^{p,p-1}$}
\label{ypp-1fig}
\end{minipage} 
\end{figure}

\begin{figure}[!h]
\centering
\includegraphics[scale=0.85]{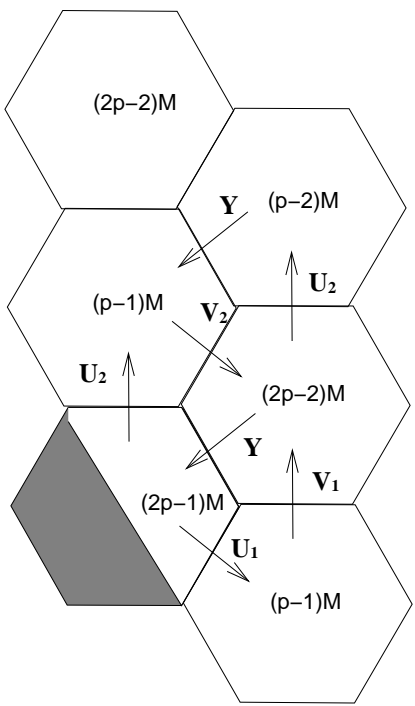}
\caption{``Bottom'' part of the dimer for  $Y^{p,p-1}$. }
\label{ypp-1-bfig}
\end{figure}

We first make a distinction among the models under scrutiny according to whether the top divided hexagon generates an ADS superpotential or not. We then focus first our attention to the case in which the top-right half of the last cut hexagon is such that $N_f<N_c$. By equation (\ref{eqADS}) imposing this condition is equivalent to
\begin{equation}
p-q < \frac{p}{q}
\label{eqADSbar}
\end{equation}
For $p$, $q$ integers, $0<q<p$ we are left with only two possibilities: $Y^{p,p-1}$ and $Y^{p,1}$. As can be seen from figure \ref{yp1fig}-\ref{ypp-1fig} for $Y^{p,1}$ we are dealing with a dimer diagram consisting of two uncut hexagons at the top of $p-1$ cut hexagons while $Y^{p,p-1}$ is described by only one cut hexagon surmounted by a tower of $2p-2$ undivided ones. In both cases each cut hexagon is such that $N_f<N_c$, i.e. develops non-perturbative terms of the ADS form.
\begin{figure}[!h]
\centering
\includegraphics[scale=0.85]{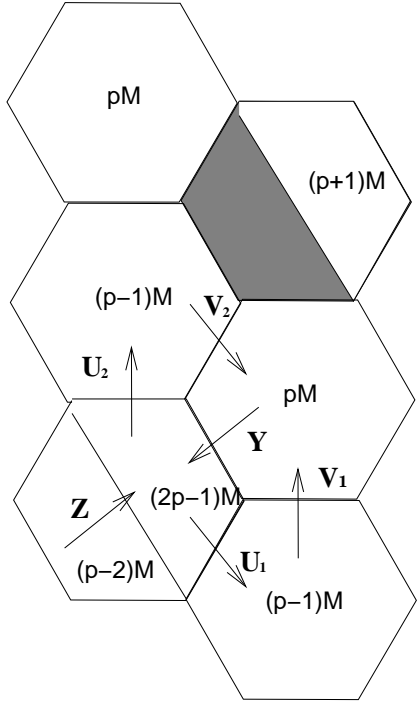}
\caption{``Top'' part of the dimer for  $Y^{p,1}$. Different hexagons are labeled (uniquely) with the rank of the associated gauge group.}
\label{yp1-bfig}
\end{figure}
We suggest that in both cases a classically flat direction described by a baryonic gauge invariant field is pushed to infinity in the quantum theory because of the appearance of the above mentioned Affleck-Dine-Seiberg terms.\\
The $Y^{p,1}$ case is most easily worked out: let us build the matrix $\mathcal{M}$ of the meson fields for the top hexagon according to the following definitions (see figure \ref{yp1-bfig}), in which for the sake of notational clarity we put the number of fractional branes $M=1$:
\bea
M_{p,p-1} & \equiv & Y_{p,2p-1}U^{(2)}_{2p-1,p-1} \\
N_{p,p-1} & \equiv & Y_{p,2p-1}U^{(1)}_{2p-1,p-1} \\
L_{p-2,p-1} & \equiv & Z_{p-2,2p-1}U^{(2)}_{2p-1,p-1} \\
O_{p-2,p-1} & \equiv & Z_{p-2,2p-1} U^{(1)}_{2p-1,p-1}
\eea
Each $X_{N_a, N_b}$ field, carrying gauge indices $a$ and $b$ in the fundamental of $SU(N_a)$ and in the antifundamental of $SU(N_b)$ respectively is actually a $N_a \times N_b$ rectangular matrix $X^{ab}_{N_a, N_b}$, $a=1 \ldots N_a$, $b=1 \ldots N_b$. Contractions of indices in the previous equations are then made in the obvious way, that is, according to the usual matrix product. The meson matrix is then:
\beq
\mathcal{M}\equiv
\left(\begin{array}{c|c}
M_{p,p-1} & N_{p,p-1} \\
\hline
L_{p-2,p-1} & O_{p-2,p-1}
\end{array}\right)
\label{defMyp1}
\eeq
In (\ref{defMyp1}) the upper blocks are $p\times (p-1)$ matrices, while the lower ones are $(p-2)\times (p-1)$ matrices; altogether, $\mathcal{M}$ is a $(2p-2)\times (2p-2)$ square matrix. The $F$-term equation for the field $V^{(2)}_{p-1,p}$ (see figure \ref{yp1-bfig}) reads\footnote{Remember that every vertex in the tiling gives a term in the tree level superpotential which is a monomial in the chiral superfield linked up with the node and with plus or minus sign alternating in such a way that every field appears inside only two monomials and with different sign.}
\begin{equation}
M_{p,p-1}=0
\label{eqMYp1}
\end{equation}
as a matrix equation, i.e. the upper-left block in $\mathcal{M}$ is identically zero due to the fact that, after decoupling of the first bottom-left quadrilateral, the $V^{(2)}_{p-1,p}$ field appears only once inside a term of the form $M_{p,p-1}V^{(2)}_{p-1,p}$ in the superpotential. Consequently 
\beq
\hbox{rank }\mathcal{M}\leq 2p-3
\eeq
which forces 
\beq
\det \mathcal{M}=0
\eeq
This in turn implies that
\bea
\label{runawayyp1-a}
\frac{\partial W_{eff}}{\partial M_{p,p-1}} = 0 & \Rightarrow & V^{(2)}_{p-1,p}= \Lambda^{4p-1}\frac{\hbox{Minor}(M_{p,p-1})^T}{\left(\det \mathcal{M}\right)^2} \rightarrow \infty \\
\frac{\partial W_{eff}}{\partial N_{p,p-1}} = 0 & \Rightarrow & V^{(1)}_{p-1,p}= -\Lambda^{4p-1}\frac{\hbox{Minor}(N_{p,p-1})^T}{\left(\det \mathcal{M}\right)^2} \rightarrow \infty
\label{runawayyp1-b}
\eea
Now we observe that we can build the following field from the $(p-1)$-fold antisymmetrized product of $(V_{p-1, p})^{\alpha}_{i,\bar j}$ fields, with $\alpha$ an $SU(2)$ replica index and $i$, $\bar j$ running on the fundamental of $SU(p-1)$ and the antifundamental of $SU(p)$ respectively,
\beq
B^{\xi}_{\bar j} \equiv \epsilon_{\bar j \bar j_1 \ldots \bar j_{p-1}} \epsilon_{ i_1, \ldots,  i_{p-1}} (V_{p-1, p})^{\alpha_1}_{ i_1 \bar j_1} \ldots (V_{p-1, p})^{\alpha_{p-1}}_{ i_{p-1} \bar j_{p-1}} 
\eeq
where $\xi \simeq \{ \alpha_1 \ldots \alpha_{p-1} \}$ is an index in the $(p-1)$-fold symmetric representation of $SU(2)$, which is a weight $(p-1)/2$ representation. Since both $\xi$ and $\bar j$ run from $1$ to $p$, we can construct a baryon out of the runaway fields $\left(V_{p-1, p}\right)^\alpha_{ i \bar j}$ in the following way
\beq
\mathcal{B} \equiv \epsilon_{\bar k_1 \ldots \bar k_{p}} \epsilon_{\xi_1, \ldots, \xi_p} B^{\xi_1}_{\bar k_1} \ldots  B^{\xi_p}_{\bar k_p}
\eeq
Here and in the following development we will always suppose that it is possible to send to infinity in an independent way the runaway fields with which we construct the baryonic direction. By (\ref{runawayyp1-a})-(\ref{runawayyp1-b}) we then argue that the $F$-terms force
\beq
\langle \mathcal{B} \rangle \rightarrow \infty
\eeq
We notice that this direction is classically flat.\\
If we remember that the gauge group factor associated to the top-right half of every cut hexagon develops an $ADS$-like term we can derive the following chain of equations relating the meson fields of each divided hexagon to the ones of the cut hexagons below and above:
\begin{eqnarray}
M^i & \propto & \frac{\hbox{Min}(L^{i+1})}{\det\mathcal{M}^{i+1}} \nonumber \\
N^i & \propto & \frac{\hbox{Min}(O^{i+1})}{\det\mathcal{M}^{i+1}} \nonumber \\
L^i & \propto & \frac{\hbox{Min}(M^{i-1})}{\det\mathcal{M}^{i-1}} \nonumber \\
O^i & \propto & \frac{\hbox{Min}(N^{i-1})}{\det\mathcal{M}^{i-1}}
\label{mesonsyp1}
\end{eqnarray}
If we suppose that it is consistent to send the fields to infinity in an independent way, since $\det\mathcal{M}=0$ for the top divided hexagon, the equations (\ref{mesonsyp1}) may suggest the following picture: the meson fields belonging to the gauge factors below get an infinite or a zero vev, according to whether they are an odd or even number of places away from the top cut hexagon; namely the second cut hexagon (counting from the top) has then $\det\mathcal{M}^2 \rightarrow \infty$, the third $\det\mathcal{M}^3 \rightarrow 0$, and so on. Clearly the real situation may be different, the important point being that we have found at least one runaway direction:  $\mathcal{B}$. \\
Let us then pass to the case $Y^{p,p-1}$ (see figure \ref{ypp-1-bfig}). In this case we have only one divided hexagon, whose bottom-left half decouples at the end of the cascade. Because of this, the structure of its meson matrix simplifies considerably:
\beq
\mathcal{M}\equiv
\left(\begin{array}{c|c}
M_{2p-2,p-1} & N_{2p-2,p-1} \\
\end{array}\right)
\label{defMypp-11}
\eeq
as we have no analog for the $O$ and $L$ mesons appearing in the previous case. Considering the equation of motion for the field $V^{(2)}_{p-1, 2p-2}$ we get that
\beq
\frac{\partial W_{eff}}{\partial V^{(2)}_{p-1, 2p-2}} = 0 \Rightarrow M_{2p-2,p-1}= U^{(2)}_{2p-2,p-2} Y_{p-2,p-1}
\eeq
Being $M_{2p-2,p-1}$ equal to a product of a $(2p-2) \times (p-2)$ matrix by a $(p-2) \times (p-1)$ matrix forces its rank to be less or equal than $p-2$:\beq
\hbox{rank} M_{2p-2,p-1} \leq p-2
\eeq
so that the rank of the whole square matrix $\mathcal{M}$ cannot be greater than $2p-3$. Again this leads to $\det \mathcal{M}=0$ and, via the $F$-term equation for $M_{2p-2,p-1}$ and $N_{2p-2,p-1}$
\bea
V^{(2)}_{p-1,2p-2} &=& \Lambda^{4p-1}\frac{\hbox{Minor}(M_{2p-2,p-1})^T}{\left(\det \mathcal{M}\right)^2} \rightarrow \infty \\
V^{(1)}_{p-1,2p-2} &=& -\Lambda^{4p-1}\frac{\hbox{Minor}(N_{2p-2,p-1})^T}{\left(\det \mathcal{M}\right)^2} \rightarrow \infty
\eea
What is still left to show is that also in this case we have runaway on the baryonic branch of the moduli space of vacua. To this purpose we see that we can easily construct a baryon out of $(V_{p-1, 2p-2})^\alpha_{ j \bar i}$ as
\bea
\mathcal{B} & \equiv & \epsilon_{\bar i_1 \ldots \bar i_{2(p-1)}}\epsilon_{ j_1 \ldots  j_{p-1}}\epsilon_{ k_1 \ldots  k_{p-1}} (V_{p-1, 2p-2})^{(1)}_{j_1 \bar i_1} \ldots (V_{p-1, 2p-2})^{(1)}_{j_{p-1} \bar i_{p-1}} \nonumber \\ & \times &(V_{p-1, 2p-2})^{(2)}_{k_1 \bar i_p}\ldots (V_{p-1, 2p-2})^{(2)}_{k_{p-1} \bar i_{2(p-1)}}
\label{baryonypp-1}
\eea
and again we see that non-perturbative effects lift a baryonic classically flat direction leaving no supersymmetric vacuum for finite values of the moduli. 

\subsection{$p-q \geq \frac{p}{q}$}

\begin{figure}[!h]
\centering
\includegraphics[scale=0.8]{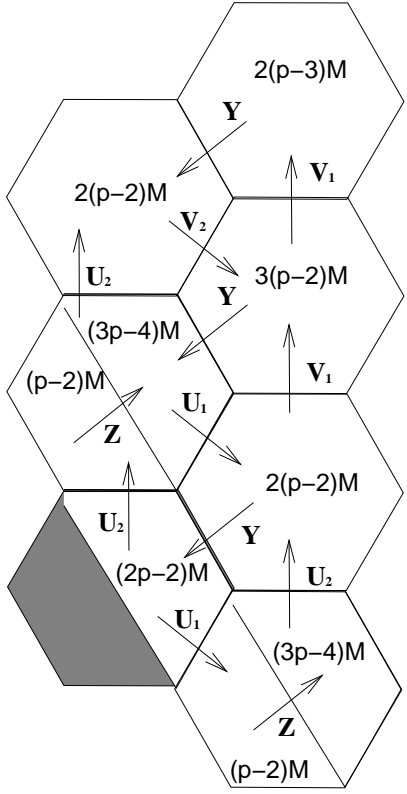}
\caption{``Bottom'' part of the dimer for  $Y^{p,p-2}$. }
\label{ypp-2-b}
\end{figure}

\begin{figure}[!t]
\centering
\includegraphics[scale=0.6]{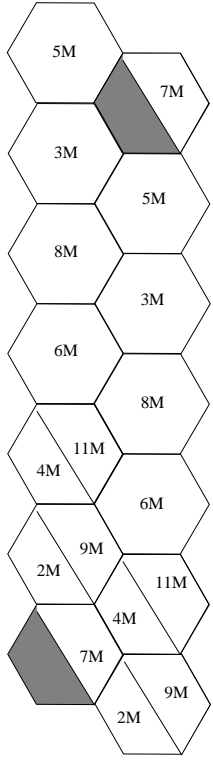}
\caption{The dimer for  $Y^{5,2}$. }
\label{y52-b-fig}
\end{figure}

The situation for $p-q \geq \frac{p}{q}$ is somewhat harder to describe but it is nevertheless possible to establish facts on the IR regime of the theory in quite complete generality. We first consider the case $q=p-2$. Here we have to deal with two divided hexagons, out of which only the lower one (see figure \ref{ypp-2-b}) generates an ADS superpotential. 
It is possible to show that in this case we still have runaway on a baryonic classically flat direction. To see that this actually happens, let us define as before
\bea
M_{2p-4,p-2} & \equiv & Y_{2p-4,2p-2}U^{(2)}_{2p-2,p-2} \\
N_{2p-4,p-2} & \equiv & Y_{2p-4,2p-2}U^{(1)}_{2p-2,p-2}
\eea
We can again construct the meson matrix $\mathcal{M}$ as in (\ref{defMypp-11}); exploiting the $SU(2p-4)\times SU(p-2)$ gauge symmetry it is possible to bi-unitarily  pseudo-diagonalize the block $M$ inside $\mathcal{M}$ so that
\beq
\mathcal{M}\equiv
\left(\begin{array}{ccccc|ccccc}
m_{11} & 0 & \dots & 0 & 0  & n_{11} & n_{12} & \dots & n_{1, p-3} & n_{1, p-2} \\
0 & m_{22} & 0 & \dots & 0   & n_{21} & n_{22} & \dots & n_{2, p-3} & n_{2, p-2} \\
0 & 0 & m_{33} & \dots & 0  & n_{31} & n_{32} & \dots & n_{3, p-3} & n_{3, p-2} \\
\vdots & \vdots & \ddots & \ddots & \vdots & \vdots & \vdots & \ddots & \ddots & \vdots \\
0 & 0 & \dots & 0 & m_{p-2, p-2} & n_{p-2,1} & n_{p-2,2} & \dots & n_{p-2, p-3} & n_{p-2, p-2} \\
0 & 0 & \dots & 0 & 0  & n_{p-1,1} & n_{p-1,2} & \dots & n_{p-1, p-3} & n_{p-1, p-2}\\
\vdots & \vdots & \ddots & \ddots & \vdots & \vdots & \vdots & \ddots & \ddots & \vdots \\
0 & 0 & \dots & 0 & 0 & n_{2p-4,1} & n_{2p-4,2} & \dots & n_{2p-4, p-3} & n_{2p-4, p-2} \\
\end{array}\right)
\label{defMypp-2}
\eeq
and consider the following set of $F$-term equations
\bea
\label{ypp2run1}
\frac{\partial W_{eff}}{\partial M_{2p-4, p-2}} = 0 & \Rightarrow & Z_{p-2, 3p-4}U^{(1)}_{3p-4, 2p-4} = \Lambda^{4p-2}\frac{ \hbox{Minor}(M_{2p-4,p-2})^T}{\left(\det \mathcal{M}\right)^{3/2}}\\
\label{ypp2run2}
\frac{\partial W_{eff}}{\partial N_{2p-4, p-2}} = 0 & \Rightarrow & Z_{p-2, 3p-4}U^{(2)}_{3p-4, 2p-4} = -\Lambda^{4p-2}\frac{\hbox{Minor}(N_{2p-4,p-2})^T}{\left(\det \mathcal{M}\right)^{3/2}}\\
\label{ypp2run3}
\frac{\partial W_{eff}}{\partial U^{(1)}_{3p-4, 2p-4}} = 0 & \Rightarrow & M_{2p-4, p-2}Z_{p-2, 3p-4} = V^{(1)}_{2p-4, 3p-6}Y_{3p-6, 3p-4} \\
\label{ypp2run4}
\frac{\partial W_{eff}}{\partial U^{(2)}_{3p-4, 2p-4}} = 0 & \Rightarrow & N_{2p-4, p-2}Z_{p-2, 3p-4} = V^{(2)}_{2p-4, 3p-6}Y_{3p-6, 3p-4} \\
\label{ypp2run5}
\frac{\partial W_{eff}}{\partial V^{(\alpha)}_{2p-4, 3p-6}} = 0 & \Rightarrow & Y_{3p-6, 3p-4}U^{(\alpha)}_{3p-4, 2p-4} = U^{(\alpha)}_{3p-6, 2p-6}Y_{2p-6, 2p-4} \\
\label{ypp2run6}
\frac{\partial W_{eff}}{\partial Y_{2p-6, 2p-4}} = 0 & \Rightarrow & V^{(1)}_{2p-4, 3p-6}U^{(1)}_{3p-6, 2p-6} = V^{(2)}_{2p-4, 3p-6}U^{(2)}_{3p-6, 2p-6} 
\eea
Multiplying from the left (\ref{ypp2run1}) by $M_{2p-4, p-2}$ and (\ref{ypp2run2}) by $N_{2p-4, p-2}$ and using (\ref{ypp2run3})-(\ref{ypp2run6}) we get
\beq
M_{2p-4,p-2}\hbox{Min}(M_{2p-4,p-2})^T \propto N_{2p-4,p-2}\hbox{Min}(N_{2p-4,p-2})^T
\label{ypp2run7}
\eeq
but since
\beq
M_{2p-4,p-2}\hbox{Min}(M_{2p-4,p-2})^T  =  \det \mathcal{M}
\left(
\begin{array}{cccc|cccc}
1 & 0  & \dots  & 0 & * & * & \dots & * \\
0 & 1 & \dots  & 0 & * & * & \dots & * \\
\vdots & \ddots & \ddots & \vdots & \vdots & \ddots & \ddots & \vdots\\
0 & 0 & \dots & 1 & * & * & \dots & * \\
\hline
0 & 0 & \dots & 0 & 0 & 0 & \dots & 0 \\
0 & 0 & \dots & 0 & 0 & 0 & \dots & 0 \\
\vdots & \ddots & \ddots & \vdots & \vdots & \ddots & \ddots & \vdots\\
0 & 0 & \dots & 0 & 0 & 0 & \dots & 0 \\
\end{array}
\right)
\eeq
\beq
N_{2p-4,p-2}\hbox{Min}(N_{2p-4,p-2})^T  =  \det \mathcal{M}
\left(
\begin{array}{cccc|cccc}
0 & 0  & \dots  & 0 & * & * & \dots & * \\
0 & 0 & \dots  & 0 & * & * & \dots & * \\
\vdots & \ddots & \ddots & \vdots & \vdots & \ddots & \ddots & \vdots\\
0 & 0 & \dots & 0 & * & * & \dots & * \\
\hline
0 & 0 & \dots & 0 & 1 & 0 & \dots & 0 \\
0 & 0 & \dots & 0 & 0 & 1 & \dots & 0 \\
\vdots & \ddots & \ddots & \vdots & \vdots & \ddots & \ddots & \vdots\\
0 & 0 & \dots & 0 & 0 & 0 & \dots & 1 \\
\end{array}
\right)
\eeq
we see that (\ref{ypp2run7}) is satisfied if and only if $\det \mathcal{M} = 0$. Therefore, defining
\bea
A^{(1)} & = & Z_{p-2, 3p-4}U^{(1)}_{3p-4, 2p-4}\\
A^{(2)} & = & Z_{p-2, 3p-4}U^{(2)}_{3p-4, 2p-4}
\eea
we can construct a baryon field in exactly the same way as in (\ref{baryonypp-1})
\bea
\mathcal{B} & \equiv & \epsilon_{\bar i_1 \ldots \bar i_{2p-4}}  \epsilon_{ j_1 \ldots  j_{p-2}} 
\epsilon_{ k_1 \ldots  k_{p-2}} A^{(1)}_{j_1  \bar i_1} \ldots A^{(1)}_{ j_{p-2} \bar i_{p-2}} \nonumber \\ & \times & A^{(2)}_{k_1 \bar i_{p-1}} \ldots A^{(2)}_{k_{p-2} \bar i_{2p-4}}
\eea
and from (\ref{ypp2run1})-(\ref{ypp2run2}) we can again conclude that the baryonic direction should display a runaway behavior
\beq
\langle \mathcal{B} \rangle \rightarrow \infty
\eeq
The general case presents many more complications due to the fact that more than one divided hexagon may generate an ADS superpotential and also because we have much less control on the expression of the meson matrix, which usually cannot be cast in the simple form (\ref{defMypp-2}). Actually, the meson matrix for a cut hexagon whose bottom-left quadrilateral does not decouple is made up of four independent blocks as in (\ref{defMyp1}):
\beq
\mathcal{M}\equiv
\left(\begin{array}{c|c}
M & N \\
\hline
O & P
\end{array}\right)
\label{defMypgen}
\eeq
The gauge freedom allows us to (pseudo)-diagonalize at most one of the
blocks in $(\ref{defMypgen})$, while no additional constraint can be put
on the a priori form of $\mathcal{M}$. As a consequence, verifying the
incompatibility of $\det \mathcal{M} \neq 0$ with the
extremization of $W_{eff}$, which is a sufficient condition to exclude a
supersymmetric vacuum for finite values of the fields, becomes a more
tricky point. We can anyway provide arguments showing
that still the theory exhibits a runaway behavior along a classically
flat direction. To this aim we will study the paradigmatic case of
$Y^{5,2}$: here we are in the quite general situation in which we must deal with three cut hexagons, two of which
generating an ADS term inside $W_{eff}$ (see figure \ref{y52-b-fig}).
As before, let us label the fields according to the rank of the gauge
groups they transform under and define also
\bea
M^{(1)}_{42} & \equiv & Y_{47}U^{(2)}_{72} \\
N^{(1)}_{42} & \equiv & Y_{47}U^{(1)}_{72} \\
M^{(2)}_{64} & \equiv & Y_{69}U^{(2)}_{94} \\
N^{(2)}_{64} & \equiv & Y_{69}U^{(1)}_{94} \\
L^{(2)}_{24} & \equiv & Z_{29}U^{(2)}_{94} \\
O^{(2)}_{24} & \equiv & Z_{29}U^{(1)}_{94}
\eea
\beq
\mathcal{M}^{(1)}\equiv
\left(\begin{array}{c|c}
M^{(1)}_{42} & N^{(1)}_{42}
\end{array}\right)
\label{defM1y52}
\eeq
\beq
\mathcal{M}^{(2)}\equiv
\left(\begin{array}{c|c}
M^{(2)}_{64} & N^{(2)}_{64} \\
\hline
L^{(2)}_{24} & O^{(2)}_{24}
\end{array}\right)
\label{defM2y52}
\eeq
The quantum superpotential, taking into account the ADS
terms from the 9 and 11 quadrilaterals, reads
\bea
W_{eff} & = & M^{(1)}_{42}L^{(2)}_{24} - N^{(1)}_{42}O^{(2)}_{24} -
N^{(2)}_{64}Z_{411}U^{(2)}_{116}  +  M^{(2)}_{64}Z_{411}U^{(1)}_{116}
\nonumber \\
& - & Y_{811}U^{(1)}_{116}V^{(1)}_{68} + Y_{811}U^{(2)}_{116}V^{(2)}_{68} -
Y_{36}V^{(2)}_{68}U^{(2)}_{83} \nonumber \\ & + & Y_{36}V^{(1)}_{68}U^{(1)}_{83} -
Y_{58}U^{(1)}_{83}V^{(1)}_{35} +  Y_{58}U^{(2)}_{83}V^{(2)}_{35}
\nonumber \\
& + & \frac{3\Lambda_1^{17}}{\left(\det \mathcal{M}^{(1)}\right)^{1/3}} + \frac{\Lambda_2^{19}}{\det \mathcal{M}^{(2)}}
\label{Weffy52}
\eea
Extremization of $W_{eff}$ then gives

\begin{figure}[!h]
\centering
\includegraphics[scale=0.75]{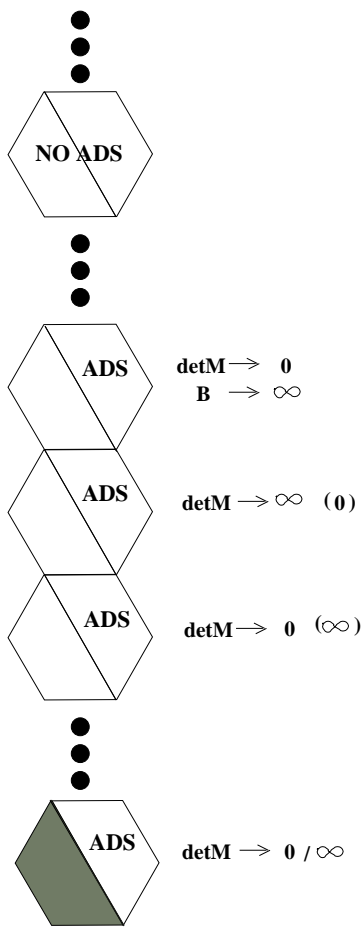}
\caption{A possible scenario for the general runaway behavior for $Y^{p,q}$, as suggested by the analysis of $Y^{5,2}$.}
\label{figygen}
\end{figure}

\begin{minipage}[!h]{0.49\linewidth}

\flushleft

\bea
L^{(2)}_{24} & = & \Lambda^{17}_1 \frac{\hbox{Min}(M^{(1)}_{42})^T}{\left( \det
  \mathcal{M}^{(1)}\right)^{4/3}}  \\
M^{(1)}_{42} & = & \Lambda^{19}_2 \frac{\hbox{Min}(L^{(2)}_{24})^T}{\left( \det
  \mathcal{M}^{(2)}\right)^2} \\
-O^{(2)}_{24} & = & \Lambda^{17}_1 \frac{\hbox{Min}(N^{(2)}_{42})^T}{\left( \det
  \mathcal{M}^{(1)}\right)^{4/3}} 
\eea
\bea
U^{(2)}_{116}N^{(2)}_{64} & = & U^{(1)}_{116}M^{(2)}_{64} \\
\label{NZXY}
N^{(2)}_{64}Z_{411} & = &  V^{(2)}_{68}Y_{811} \\
\label{MZXY}
M^{(2)}_{64}Z_{411} & = &  V^{(1)}_{68}Y_{811} \\
U^{(1)}_{116}V^{(1)}_{68} & = & U^{(2)}_{116}V^{(2)}_{68} \\
\label{YXXY2}
Y_{811}U^{(2)}_{116} & = &  U^{(2)}_{83}Y_{36} \\
\label{YXXY}
Y_{811}U^{(1)}_{116} & = &  U^{(1)}_{83}Y_{36}
\eea

\vspace{.5cm}

\end{minipage}
\begin{minipage}[!h]{0.54\linewidth}

\flushleft

\addtolength{\hoffset}{-.5cm}
\bea
N^{(1)}_{42} & = & -\Lambda^{19}_2
\frac{\hbox{Min}(O^{(2)}_{24})^T}{\left( \det
  \mathcal{M}^{(2)}\right)^2} \\
\label{ZWN}
Z_{411}U^{(2)}_{116} & = & -\Lambda^{19}_2
\frac{\hbox{Min}(N^{(2)}_{64})^T}{\left( \det
  \mathcal{M}^{(2)}\right)^2} \\
Z_{411}U^{(1)}_{116} & = & \Lambda^{19}_2
\frac{\hbox{Min}(M^{(2)}_{64})^T}{\left( \det
  \mathcal{M}^{(2)}\right)^2}
\label{ZWM}
\eea
\bea
\label{YWWY}
V^{(2)}_{68}U^{(2)}_{83} & = & V^{(1)}_{68}U^{(1)}_{83} \\
\label{YVVY}
Y_{36}V^{(2)}_{68} & = & V^{(2)}_{35}Y_{58} \\
Y_{36}V^{(1)}_{68} & = & V^{(1)}_{35}Y_{58} \\
U^{(2)}_{83}V^{(2)}_{35} & = & U^{(1)}_{83}V^{(1)}_{35} \\
\label{YUzero}
Y_{58}U^{(1)}_{83} & = & 0 \\
Y_{58}U^{(2)}_{83} & = & 0
\eea

\vspace{.5cm}
\end{minipage}
To see the implications of the above system of equations, we use the gauge freedom to decompose $M^{(2)}_{64}$ into singular
values, so that the $8\times 8$ matrix $\mathcal{M}^{(2)}$
(\ref{defM2y52}) has the following form
\beq
\mathcal{M}^{(2)}=\left(
\begin{array}{cccc|cccc}
m_{11} & 0 & 0 & 0 & n_{11} & n_{12} & n_{13} & n_{14} \\
0 & m_{22} & 0 & 0 & n_{21} & n_{22} & n_{23} & n_{24} \\
0 & 0 & m_{33} & 0 & n_{31} & n_{32} & n_{33} & n_{34} \\0 & 0 & 0 & m_{44} & n_{41} & n_{42} & n_{43} & n_{44} \\
0 & 0 & 0 & 0 & n_{51} & n_{52} & n_{53} & n_{54} \\
0 & 0 & 0 & 0 & n_{61} & n_{62} & n_{63} & n_{64} \\
\hline
l_{11} & l_{12} & l_{13} & l_{14} & o_{11} & o_{12} & o_{13} & o_{14} \\
l_{21} & l_{22} & l_{23} & l_{24} & o_{21} & o_{22} & o_{23} & o_{24}
\end{array}
\right) 
\label{My52-2}
\eeq
Multiplying from the left (\ref{ZWM}) by $M^{(2)}_{64}$ and using
(\ref{MZXY}), (\ref{YXXY}) yields
\beq
\Lambda_2^{19} M^{(2)}_{64}\frac{\hbox{Min}(M^{(2)}_{64})^T}{\left( \det
  \mathcal{M}^{(2)}\right)^2} =  V^{(1)}_{68}U^{(1)}_{83}Y_{36}
\label{MY36}
\eeq
The same reasoning, multiplying now (\ref{ZWN}) by $N^{(2)}_{64}$
and exploiting the relations (\ref{NZXY}),(\ref{YXXY2}) and (\ref{YWWY}) eventually brings us to 
\beq
-\Lambda_2^{19} N^{(2)}_{64}\frac{\hbox{Min}(N^{(2)}_{64})^T}{\left( \det
  \mathcal{M}^{(2)}\right)^2} =  V^{(1)}_{68}U^{(1)}_{83}Y_{36}
\label{NY36}
\eeq
so that
\beq
M^{(2)}_{64}\hbox{Min}(M^{(2)}_{64})^T =
-N^{(2)}_{64}\hbox{Min}(N^{(2)}_{64})^T
\label{MN64}
\eeq
That $(\ref{MN64})$ actually implies $\det \mathcal{M}^{(2)}=0$ can be seen
by a direct computation in Mathematica. By the way we can provide
many arguments for this to happen which may be extended to the general case.
By (\ref{MY36}) we have that the product
$M^{(2)}_{64}\hbox{Min}(M^{(2)}_{64})^T$ has at most rank equal to the
rank of $Y_{36}$ ($\leq 3$). This means that either $M^{(2)}_{64}$ or
its cofactor matrix $\hbox{Min}(M^{(2)}_{64})$ has rank  $\leq 3$. If $M_{64}$
has rank less than four we have
\bea
M^{(2)}_{64}\hbox{Min}(M^{(2)}_{64})^T & = & \left(
\begin{array}{ccc|ccc}
* & * & * & * & * & * \cr
* & * & * & * & * & * \cr
* & * & * & * & * & * \cr
\hline
0 & 0 & 0 & 0 & 0 & 0 \\
0 & 0 & 0 & 0 & 0 & 0 \\
0 & 0 & 0 & 0 & 0 & 0 \\
\end{array}
\right)
\\
N^{(2)}_{64}\hbox{Min}(N^{(2)}_{64})^T & = & \left(
\begin{array}{ccc|ccc}
* & * & * & * & * & * \cr
* & * & * & * & * & * \cr
* & * & * & * & * & * \cr
\hline
0 & 0 & 0 & 1 & 0 & 0 \\
0 & 0 & 0 & 0 & 1 & 0 \\
0 & 0 & 0 & 0 & 0 & 1 \\
\end{array}
\right)
\det\mathcal{M}^{(2)}
\eea
and thus $\det\mathcal{M}^{(2)}=0$. Suppose then that $\hbox{rank} (M_{64}^{(2)})
= 4$, and so $\hbox{rank} (\hbox{Min}M_{64}^{(2)}) = \hbox{rank} (Y_{36})$. In the generic case $\hbox{rank} (Y_{36}M_{64}^{(2)})=3$ we have that, by (\ref{YVVY}) and (\ref{YUzero}), the six 4-dimensional column vectors of $\hbox{Min}(M_{64}^{(2)})^T$ are in the kernel of a rank three operator, i.e. $\hbox{Min}(M_{64}^{(2)})^T$ has rank 1, which again forces the matrix $\mathcal{M}^{(2)}$ to be singular. The remaining cases can then be more easily studied by a direct numerical analysis, yielding in any case $\det\mathcal{M}^{(2)}=0$. A baryon field $\mathcal{B}$ can then be constructed out of $Z_{411}U^\alpha_{116}$ in the same fashion as for $Y^{p,1}$. We thus suggest that, along the same line of the previous sections, the gauge theory has a classically flat baryonic runaway direction: 
\bea
\langle \mathcal{B} \rangle \rightarrow \infty 
\eea
As in section \ref{mes} we could also argue that $\langle \det \mathcal{M}^{(1)} \rangle \rightarrow \infty $, but clearly here the situation is less generic than in the case of the field $\mathcal{B}$ and needs more study. The important fact is that also in this paradigmatic case we have found at least one good candidate runaway direction ($\mathcal{B}$). Although very difficult to prove in general with a direct computation, we argue that the arguments presented here should go through for the entire remaining class of $Y^{p,q}$ dual gauge theories, and so we expect the scenario for $Y^{5,2}$ to be reproduced in the general case. First of all, in all cases we indeed have a matrix of the form $(\ref{My52-2})$ and an equation like $(\ref{MN64})$, which has been seen to cause the runaway in the examples described in this section. Moreover, we always have a low rank matrix $Y$ entering the second hexagon (starting to count from the top of the brane tiling) that plays the same role as the one of $Y_{36}$ for the $Y^{5,2}$ runaway theory. \\ The analysis we have done in these sections suggests that the general pattern for the generic gauge theory dual to a $Y^{p,q}$ manifolds is the following: the meson matrix for the first cut hexagon generating an ADS term is forced to have zero determinant by the extremization of $W_{eff}$; this will in turn cause a gauge-invariant baryonic classical flat direction to be sent to infinity. For this reason we can claim that every gauge theory dual to a $Y^{p,q}$ manifold with $p>q>0$ has no stable vacuum supersymmetric or not, but should instead develop a runaway behavior. It would be interesting to verify in all details if this is the only runaway direction or as we have suggested in \ref{mes} it is possible that the general runaway behavior is the one shown in figure \ref{figygen}. Namely, not only do we have the baryonic runaway direction $\mathcal{B}$, but also the meson matrices for the cut hexagons below the first one that develop an $ADS$ term will alternatively get values equal to zero or infinity and allow the construction of runaway gauge invariant directions by taking their determinant. 

\section{A look beyond the $Y^{p,q}$ case}
\label{BeyondYpq}

Until now we have limited ourselves to the analysis of the $C(Y^{p,q})$ family of singularities. This is clearly a very small part of all the possible singularities for which we have powerful tools to construct the dual gauge theory: indeed, using the recently discovered Fast-Inverse-Algorithm \cite{hananyrombi} we can in principle associate the gauge theories dual to every type of toric cone singularities, which gives us the possibility of a general discussion of the $IR$ dynamics for all toric singularities. We hope to address this ambitious program in a future work. In this section we only want to outline how the runaway behaviour found in the previous sections is not a particular characteristic of the $C(Y^{p,q})$ singularities, but it in fact extends to more general cases in which no more $SU(2)$ doublets appear, but we have only abelian $T^3$ symmetry. In fact as toric varieties the $C(Y^{p,q})$ are highly constrained due to the presence of the non-abelian $SU(2)$ isometry factor; for this reason it is worthwhile to take a closer look at a more general manifold as a base of the $CY$ cone\footnote{We will use the notation of \cite{zaffa}, with $L^{p,q;r,s} \rightarrow L^{a,b;c,d}$.}: $L^{a,b;c,d}$. We refer to \cite{kru2,francolpqr,zaffa} for a complete description of the dual gauge theories. We first of all point out that in this case, unless these theories can be represented by a single row of regular and cut hexagons, we do not have an explicit map that tells us the disposition of the cuts in the dimers associated to a particular geometry. This makes it hard to find a general pattern for the analysis of the IR regime. Nevertheless, still we can get some hint about the low energy behavior at least for a subclass of these theories. 
\\ From now on we will stick to the case with a single cut hexagon ($m=1$). For this class we do not have the problem of the disposition of the cut hexagons, which suggests that this subfamily could be fully analyzed. We will suppose to be at the end of a cascade of Seiberg dualities and we will discuss what happens to the theory in the far $IR$. We will not try to go through the details of the $IR$ dynamics in general, but we can anyway shed some light on how the runaway behavior found for the $Y^{p,q}$ family might actually be reproduced in the $L^{a,b;c,d}$ class. To this purpose we will first make some general considerations about the existence of runaway for a subfamily of the $m=1$ class which is particularly easy to study, and then work out in two specific cases the runaway lift of a classically flat baryonic direction. The first one bears some resemblances with the $Y^{p,q}$ gauge theories because it has some double arrows in the quivers and the form of the superpotential in the $IR$ is not the most general. The second instead is a little more general and can give us the idea of the mechanism for runaway in the subclass with $m=1$, even if we have no argument to exclude that the possible variety of dynamics may very well be much richer than the one explored in this section. Moreover, our study still leaves open the possibility that some ``compensating'' effect between ADS terms might lead to something unexpected in the case with more than one divided hexagon. \\
First of all, let us start our considerations sketching the argument for the runaway behavior of a subclass of dimer theories dual to $L^{a,b;c,d}$ characterized by $n$ regular hexagons, one cut hexagon ($m=1$) and periodicity $k=2$. This is the mildest generalization in the context of the $L^{a,b;c,d}$ theories of the $Y^{p,p-1}$ family considered before: it has only one cut hexagon but dimer periodicity two instead of one. \\
In particular we want to consider the subclass with $a = 1$, $b= 6l+1$, $c = 2l+1$ and $d = 4l+1$ where $l$ is a non-negative integer such that it satisfies the regularity conditions for the manifold. The dual gauge theories have dimer diagrams with one cut hexagon and a tower of $n$ regular hexagons, where $n=6l$ is a multiple of six.
\begin{figure}[!h]
\centering
\includegraphics[scale=0.75]{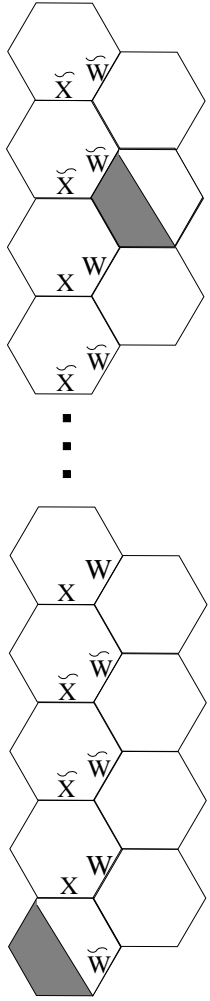}
\caption{The brane tiling for $L^{1,6l+1;2l+1,4l+1}$ with $m=1$, $k=2$.}
\label{l1qrsfig1}
\end{figure} 
\begin{figure}[!h]
\centering
\includegraphics[scale=0.5]{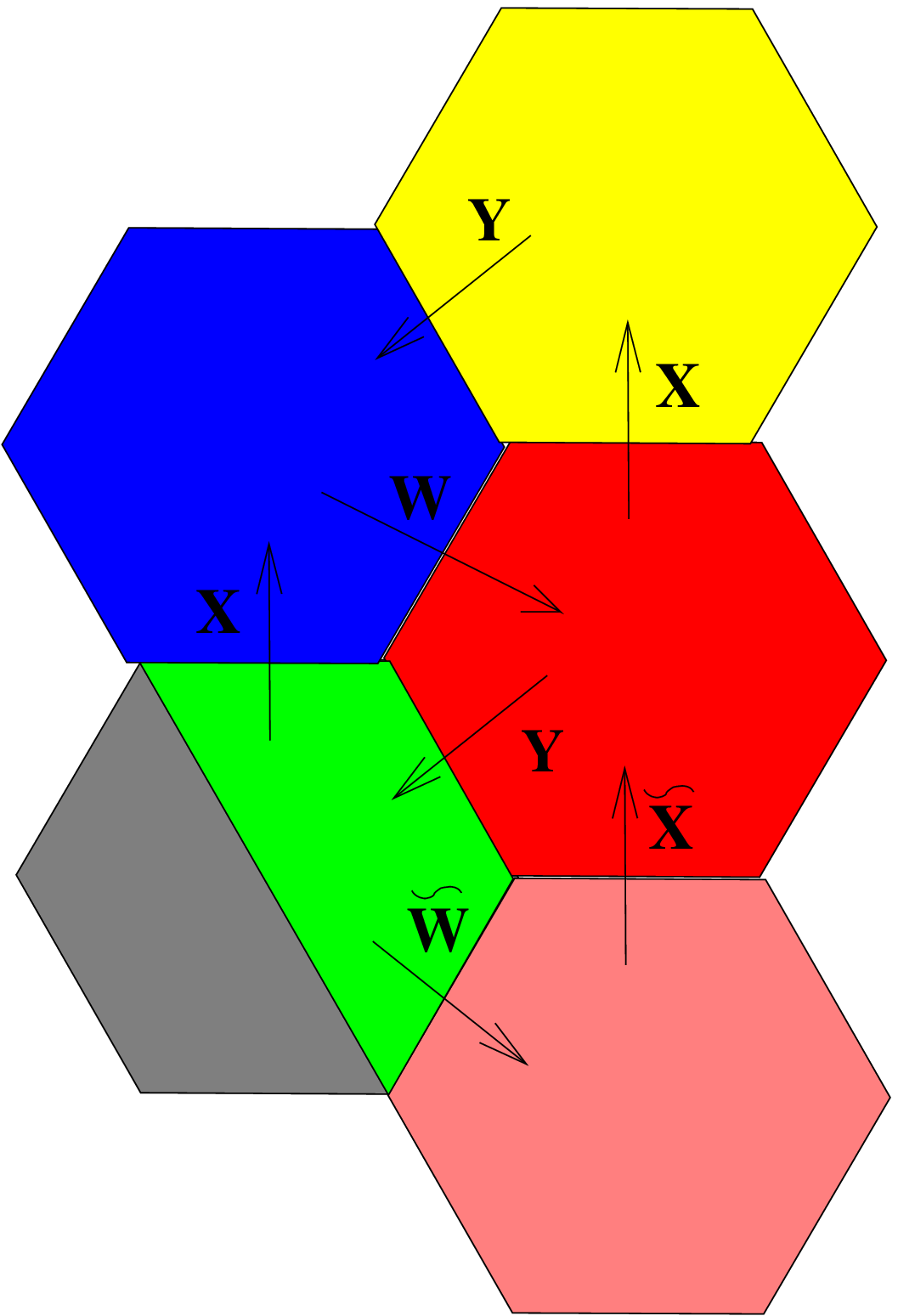}
\caption{The hexagons around the ADS quadrilateral for $L^{1,6l+1;2l+1,4l+1}$ with $m=1$, $k=2$.}
\label{l1qrsfig2}
\end{figure}

We thus have a brane tiling with one cut hexagon and a tower of $n$ regular hexagons, where $n=6l$ is a multiple of six. Consequently, the assignment of the $\tilde X$, $\tilde W$, $X$ and $W$ fields (see \cite{zaffa} for further details), due to the $k=2$ periodicity, always follows the same pattern as in figure \ref{l1qrsfig1}. We first notice that the bottom-left quadrilateral of the cut hexagon is the only decoupled gauge factor at the end of the cascade. Let us then focus on the lowest part of the diagram, referring in the following to figure \ref{l1qrsfig2}. Setting $M=1$ purely to simplify notation, the number of colors of the hexagons neighboring the non-decoupled quadrilateral are (see fig. \ref{l1qrsfig2})
\bea
\hbox{Red hexagon:} \quad N^R_C &=& 3(b-d) = 6l \\
\hbox{Blue hexagon:} \quad N^B_C &=& b-d = 2l \\
\hbox{Yellow hexagon:} \quad N^Y_C &=& 3(b-d)-d = 2l-1 \\
\hbox{Pink hexagon:} \quad N^P_C &=& 2(b-d) = 4l \\
\hbox{Green quadrilateral:} \quad N^G_C &=& b = 6l+1
\eea
Since $N^G_F=N^R_C<N^G_C$, we see that the cut hexagon contributes an ADS term to the effective superpotential. Moreover, defining the meson field $M_{RB}\equiv Y_{RG}X_{GB}$ and taking the derivative of the superpotential with respect to $W_{BR}$ we get
\beq
M_{RB}=X_{RY}Y_{YB}
\label{mesonl1qrs}
\eeq
The right hand side in (\ref{mesonl1qrs}) is a product of a $6l\times 2l-1$ by a $2l-1 \times 2l$ matrix, so it is a $6l \times 2l$ of rank (at most) $2l-1$. We can finally define the meson field $N_{RP}\equiv Y_{RG}\tilde{W}_{GP}$ and construct out of $M_{RB}$ and $N_{RP}$ a $6l \times 6l$ meson matrix
\beq
\mathcal{M}=
\left(
\begin{array}{c|c}
M_{BR} & N_{PR}
\end{array}
\right)
\eeq 
whose first $2l$ lines are linearly dependent since $\hbox{rank}(M_{BR})=2l-1$, which implies $\det\mathcal{M}=0$. Thus, for instance, the $F$-term equation for $M_{BR}$ now reads, restoring $M$ in the formulas,
\beq
W_{BR}=\frac{\Lambda^{(12l-1)M}\hbox{Minor}M_{BR}}{(\det\mathcal{M})^{\frac{1}{M}+1}} \rightarrow \infty
\eeq
This line of reasoning reveals that the same mechanism used to show the possible runaway behavior for the gauge theories dual to $Y^{p,q}$ is at work also in this subclass of this more general set of theories, and hence it seems reasonable to conclude that also these theories might display a runaway behavior. 
We will not go into the case of higher periodicity or the construction of gauge invariant fields;  let us nevertheless state that in all the examples we checked, like $L^{1,5;2,4}$ below, a very similar pattern to the one discussed above is found, leading to the same results and allowing us to conjecture that the behavior described in the previous sections is not specific to $Y^{p,q}$ and it in fact may extend beyond the cases already studied.

\subsection{Examples}
To strengthen the observations we made in the previous section we now pass to the direct study of two examples, showing besides that we are able to construct gauge invariant operators with the fields that, being coupled to the mesons of the group that develops an $ADS$ superpotential, may have a runaway behavior, plus a minimal amount of other fields. Moreover, all these fields (runaway + ``auxiliary'') are a subset of a perfect matching of the dimer \cite{dimers}. This tells us immediately that the $F$-term equations for the classical superpotential will be automatically satisfied. This happens because in our theories the superpotential has at least cubic terms (we are working at low energy and we integrate out the massive fields), so the links contained in a perfect matching will never appear in pairs in a term of the superpotential, but there will be only one for each term in $W_{tree}$. For these reasons if we turn on only a subset of links contained in a perfect matching of the theory and we put all the other vevs to zero all the $F$-term equations will be automatically satisfied.
By doing this we have a gauge invariant parametrization (solution of $D$-term equations) of a classically flat direction that is lifted by non-perturbative effects into a runaway direction. This gives us some support on the runaway behaviour of the gauge theories\footnote{A notational convention: in the following two examples the fields will not be characterized by the number of colors of the group factors they transform under, but by labels that respect the order of the gauge group factors in the tiling.}.
 
\subsection{$L^{1,5;2,4}$}

\begin{figure}
\centering
\includegraphics[scale=0.6]{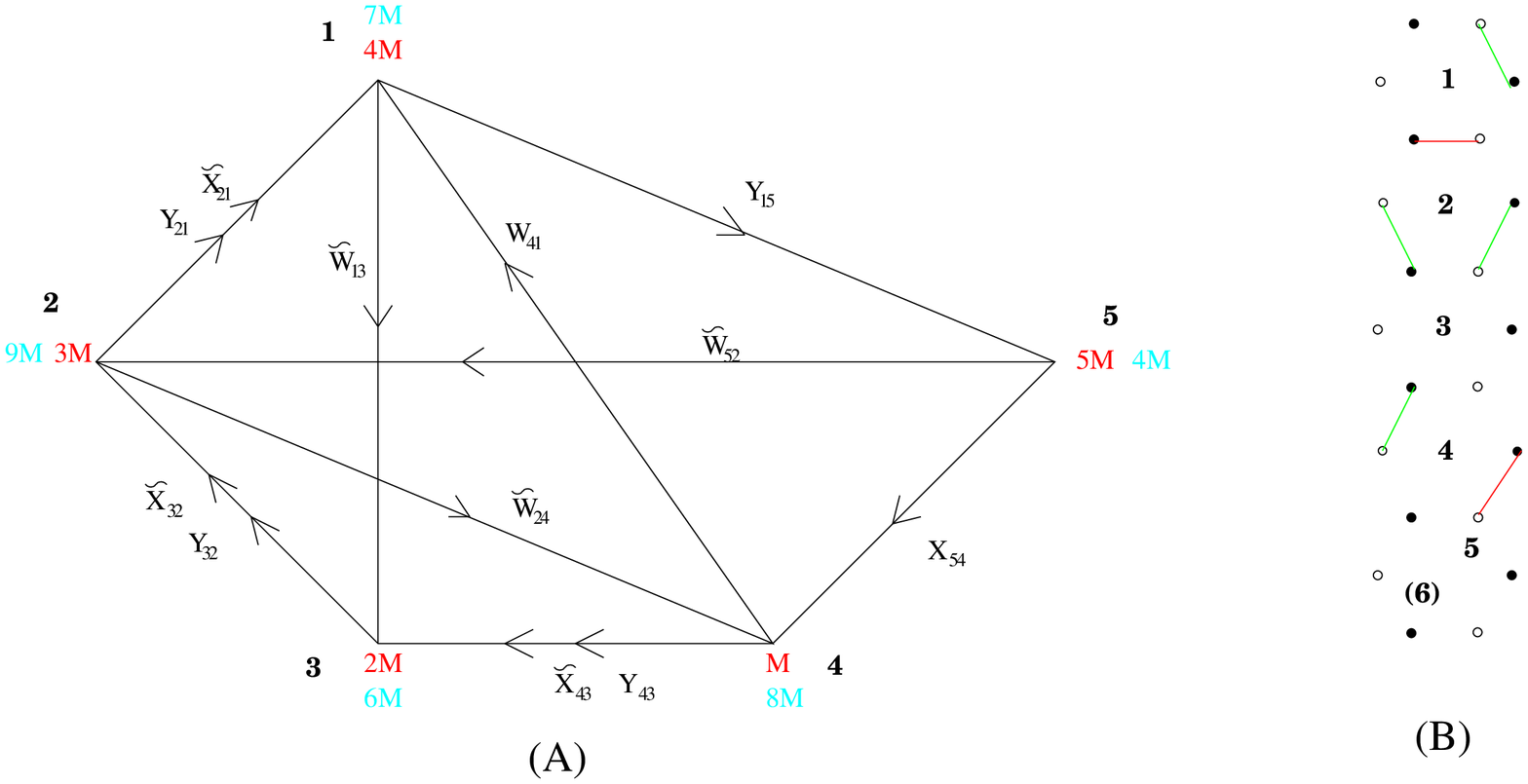}
\caption{(A) The quiver gauge theory of $L^{1,5;2,4}$ at the end of the cascade. In red the colours of the gauge groups, while in blue the number of flavours (B) The subset of the links of a perfect matching turned on to construct the gauge invariant runaway direction.}
\label{L1524-b}
\end{figure}

This theory has $m=1$ but it is outside the family discussed in the previous section. The study of the $IR$ dynamics for this theory was sketched in \cite{zaffa} and here we complete the discussion.\\
At the end of the cascade of the $RG$ flow we obtain a theory described by the quiver diagram in figure \ref{L1524-b}, which has $k=3$ shift, with a tree level superpotential given by:
\begin{equation}
\label{wtreel1524}
W_{\hbox{tree}}= -Y_{21}\tilde W_{13}\tilde X_{32} + \tilde W_{13} Y_{32}\tilde X_{21} - Y_{32}\tilde W_{24}\tilde X_{43} + \tilde W_{24} Y_{43}\tilde X_{32} +W_{41} Y_{15} X_{54} -\tilde W_{52}\tilde X_{21} Y_{15}
\end{equation}
Assuming that the $IR$ dynamics is dominated by the node with rank $5M$, it is easy to see that this one develops an $ADS$ superpotential. To describe the $IR$ dynamics we can construct the meson fields using the composite fields $Y_{15}\tilde W_{52}= M_{12}$, $Y_{15} X_{54}= N_{14}$:
\beq
\mathcal{M}\equiv
\left(\begin{array}{c|c}
M_{12} & N_{14}
\end{array}\right)
\label{mesl}
\eeq
The complete effective superpotential is given by:
\bea
\label{weffl1524}
W_{\hbox{eff}} &=& -Y_{21}\tilde W_{13}\tilde X_{32} + \tilde W_{13} Y_{32}\tilde X_{21} - Y_{32}\tilde W_{24}\tilde X_{43} + \tilde W_{24} Y_{43}\tilde X_{32} \nonumber \\ &+& W_{41} N_{14} - M_{12} X_{21} + \frac{\Lambda _5 ^{11} }{(\det \mathcal{M})^{1/M}}
\eea
The important parts are the last three terms. 
Here the situation is easy to understand: the field $W_{41}$ appears in the superpotential only coupled to $N_{14}$; its $F$-term condition puts the value of $N_{14}$ equal to zero so that $\det \mathcal{M}$ will be equal to zero and the value of $W_{41}$ will be pushed to infinity.
What remains is to show that we can construct a gauge invariant operator with a particular subset of the possible perfect matchings of the theory (fig. \ref{L1524-b}).
We will use the following fields: $Y_{21}$, $\tilde X_{21}$, $\tilde W_{24}$, $W_{41}$. We put $M=1$ for simplicity, but the construction is general.
We form the composite field $\tilde W_{24} W_{41} $ and we collect the fields in a vector:
\begin{equation}
\label{vector}
V_{\alpha}=(Y_{21},\tilde X_{21},\tilde W_{24} W_{41})
\end{equation}
where we can consider $\alpha $ as an index of $SU(3)$.
Now we can construct the following field:
\begin{equation}
\label{comp}
O^{I_4}_{\{\alpha \beta \gamma \} } = \epsilon _{I_1 I_2 I_3 I_4} \epsilon _{a_1 a_2 a_3} V_{\alpha}^{a_1 I_1}  V_{\beta}^{a_2 I_2}  V_{\gamma}^{a_3 I_3} 
\end{equation}
this transforms in the fundamental of the $SU(4)$ colour and in the threefold symmetric of the fictitious $SU(3)$. Thanks to these transformation properties we can build the following gauge invariant
\begin{equation}
\label{gauinv}
Q = \epsilon _{I_1 I_2 I_3 I_4} O^{I_1}O^{I_2}O^{I_3}O^{I_4}
\end{equation}  
The existence of this gauge invariant classical flat direction that contains the runaway elementary fields suggests that the behavior of the theory is runaway.

\subsection{$L^{1,7;3,5}$}
 
\begin{figure}[!h]
\centering
\includegraphics[scale=0.6]{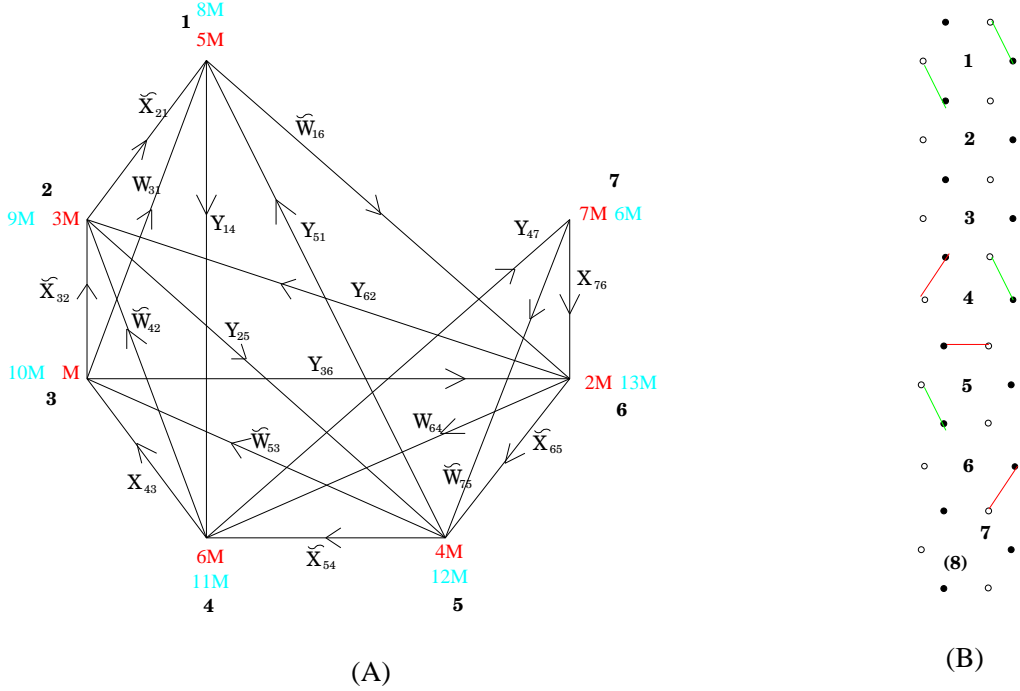}
\caption{(A) The quiver gauge theory of $L^{1,7;3,5}$ at the end of the cascade. In red the colours of the gauge groups, while in blue the number of flavours. (B) The subset of links of a perfect matching turned on to construct the gauge invariant runaway direction.}
\label{L1735}
\end{figure}

In this case the quiver diagram is the one in figure \ref{L1735}, which represents a case with $k=2$ shift. The tree level superpotential is:
\begin{eqnarray}
\label{wtreel1735}
W_{\hbox{tree}} & = & -Y_{51}\tilde W_{16}\tilde X_{65} + \tilde W_{16} Y_{62}\tilde X_{21} - Y_{14}\tilde W_{42}\tilde X_{21} + W_{31} Y_{14} X_{43} - Y_{25}\tilde W_{53} \tilde X_{32} +\nonumber \\ & + & \tilde X_{54} Y_{52} \tilde W_{42} + \tilde X_{65} \tilde W_{53} Y_{36} - W_{64} X_{43} Y_{36} + X_{76}W_{64}Y_{47} -Y_{47}\tilde W_{75} \tilde X_{54}
\end{eqnarray}
Again, let us assume that the $IR$ dynamics is dominated by the rank $7M$ node, which is characterized by $N_F - N_C = -M$. It therefore generates an $ADS$ superpotential and, as in the above mentioned case, it is the only one which does. Again, in order to describe the gauge-invariant degrees of freedom relevant for the low energy regime let us define the meson fields  $Y_{47}\tilde W_{75}\equiv N_{45}$, $Y_{47} X_{76}\equiv M_{46}$:
\beq
\mathcal{M}\equiv
\left(\begin{array}{c|c}
M_{46} & N_{45}
\end{array}\right)
\label{mesl2}
\eeq
Plugging their non-perturbative contribution to the full effective superpotential we thus get
\begin{eqnarray}
\label{weffl1735}
W_{\hbox{eff}} & = & -Y_{51}\tilde W_{16}\tilde X_{65} + \tilde W_{16} Y_{62}\tilde X_{21} - Y_{14}\tilde W_{42}\tilde X_{21} + W_{31} Y_{14} X_{43} - Y_{25}\tilde W_{53} \tilde X_{32} + \nonumber \\ & + & \tilde X_{54} Y_{52} \tilde W_{42} + \tilde X_{65} \tilde W_{53} Y_{36} - W_{64} X_{43} Y_{36} + M_{46}W_{64} - N_{45} \tilde X_{54}  + \nonumber \\ & + & \frac{\Lambda_7 ^{11} }{(\det \mathcal{M})^{1/M}}
\end{eqnarray}
Because this theory is a particular case of the class  $L^{1,6l+1;2l+1,4l+1}$ with $l=1$, the runaway mechanism is completely analogous to what we are by now familiar for the general $k=2$ case. We are then forced to have $\det\mathcal{M}=0$, causing $W_{64}$ to get an infinite vev via the $F$-term equation for $M_{46}$. The same reasoning holds for $\tilde X_{54}$ and the meson field $N_{45}$ playing the role of $W_{64}$ and $M_{46}$ respectively.
To construct a gauge invariant operator out of a subset of the perfect matchings we will consider in this case the fields $Y_{14}$, $Y_{51}$ together with the two runaway fields $\tilde X_{54}$ and $W_{64}$ (we put again $M=1$ for simplicity). We first notice that we can write down a baryonic direction using $W_{64}$ only as follows: define first 
\beq
P_{ik}\equiv\epsilon_{JL}\epsilon_{IJ}\epsilon_{KL}(W_{64})_{Ii}(W_{64})_{Kk}
\eeq
where now upper case indices run from 1 to 2, while lower case indices run from 1 to 6. We can form a baryon simply taking the threefold antisymmetric product of $P_{ik}$
\beq
\mathcal{B}_1 = \epsilon_{ijklmn}P_{ij}P_{kl}P_{mn}
\eeq
As far as the runaway direction induced by the behavior of $\tilde X_{54}$ is concerned, it can be parameterized by a gauge-invariant baryonic composite field in the following fashion. Define
\bea
\begin{array}{ccll}
Q^1_{ik} & \equiv &  (\tilde X_{54})_{ik} & \quad i=1\ldots 4, \quad k=1\ldots 6 \\
Q^2_{ik} & \equiv &  (Y_{14})_{il}(Y_{51})_{lk} & \quad i=1\ldots 4, \quad l=1 \ldots 5, \quad k=1\ldots 6 \\
\end{array}
\eea
Notice that, for generic values $Y_{14}$ and $Y_{51}$, which are $4\times5$ and $5\times6$ matrices respectively, $Q^2$ is a generic $4\times 6$ matrix and in particular it may have maximal rank, as opposed to the previously examined case of the product matrix $\tilde W_{24} W_{41}$ for $L^{1,5;2,4}$. We may thus write
\bea
\mathcal{B}_2 & = & \epsilon_{i_1\ldots i_6} \epsilon_{j_1 \ldots j_6}  Q^1_{i_1 j_1} \ldots Q^1_{i_4 j_4} Q^1_{i_5 b_1} Q^1_{i_6 b_2} Q^2_{j_1 b_3}Q^2_{j_2 b_4}Q^2_{j_3 c_1} \ldots Q^2_{j_6 c_4} \nonumber \\
& \times & \epsilon_{a_1 \ldots a_4} \epsilon_{b_1 \ldots b_4} \epsilon_{c_1 \ldots c_4}
\eea
Again, we are led to the conclusion that the runaway behavior of $\tilde X_{54}$ and $W_{64}$ turns baryonic classically flat directions into runaway.

\section{Conclusions and outlook}
\label{Conclusions}
In this paper we have analyzed the infrared behavior for a wide class of quiver gauge theories with known gravity duals. These include the whole $Y^{p,q}$ family and a subclass of $L^{a,b,c}$ singularities. We have provided arguments that the general scenario is such that a baryonic classically flat direction is lifted to infinity due to non-perturbative effects, which suggests that the complete dynamics does not allow a non-supersymmetric stable vacuum for finite values of the moduli. This is clearly a point which deserves further investigation: in particular, it would be interesting (though very difficult) to consider how the K\"ahler dynamics may affect this picture. Moreover, even if we have arguments that tend to exclude the presence of a stable non-supersymmetric vacuum for this class of theories, the question about the existence of metastable vacua is completely left open \cite{seibergmeta}.\\ The runaway behavior on the gauge theory side rises many interesting questions about what happens to the supergravity solution at the end of the throat. Everything seems to suggest some kind of instability in the full solution, but the gravity dual of a runaway behavior is not yet well understood. \\
Now that we have a definite map between toric singular Calabi-Yau cone and dual quiver gauge field theories it would be interesting to study the subject of dynamical supersymmetry breaking gauge theories with a gravity dual in the generic toric case. A particularly exciting point would be to understand if it is possible to realize, in the context of the theories allowing a dimer representation, dynamically supersymmetry breaking without runaway; this would be a remarkable achievement in the study of the gauge/gravity correspondence since it would lead, in line of principle, to the possibility to construct a string solution with a stable non-supersymmetric vacuum. 

\vskip 1truecm
 \noindent {\Large{\bf Acknowledgements}}
\\
It is a pleasure to thank Alberto Zaffaroni and Agostino Butti for many enlightening discussions and their constant support. We are also grateful to  Matteo Bertolini for many suggestions and discussions, Amihay Hanany for a kind and instructive correspondence and to Angel M. Uranga for a helpful conversation.\\
We would also like to thank Diego Gallego, Houman Safaai, Luca Vecchi, Fabio Ferrari Ruffino, Luca Philippe Mertens, Annibale Magni for useful discussions.\\ Moreover, we are sincerely grateful to the Theory Group at the University of Milano-Bicocca, where part of this work was done. \\
This research has been supported by the Italian MIUR under the program ``Teoria dei campi, superstringhe e gravit\`a'' and the European Commission RTN Program MRTN-CT-2004-005104.

\end{document}